\documentclass[aps,prd,twocolumn,showpacs,superscriptaddress,nofootinbib,10pt]{revtex4-1}

\usepackage{amsfonts}
\usepackage{amsmath}
\usepackage{amssymb}
\usepackage{bm}
\usepackage{dcolumn}
\usepackage{epsfig}
\usepackage{graphics}
\usepackage{graphicx}
\usepackage{hyperref}
\usepackage[latin1]{inputenc}
\usepackage{latexsym}
\usepackage{multirow}
\usepackage{rotating}
\usepackage{upgreek}

\newcommand{\mb}[1]{\mbox{\boldmath $#1$}}

\newcommand{\APO}{{\mbox{\tiny apo}}}
\newcommand{\INC}{{\mbox{\tiny inc}}}

\newcommand{\met}{\mbox{g}}

\newcommand{\MIN}{{\mbox{\tiny min}}}
\newcommand{\NN}{{\mbox{\tiny N}}}
\newcommand{\PERI}{{\mbox{\tiny peri}}}
\newcommand{\pont}{{\,^\ast\!}R\,R}

\newcommand{\TT}{{\mbox{\tiny TT}}}

\newcommand*{\IEEC}{Institut de Ci\`encies de l'Espai (CSIC-IEEC), Campus UAB, Torre C5 parells, 08193 Bellaterra, Spain}
\newcommand*{\IAO}{Institute of Astronomy, Madingley Road, Cambridge, CB30HA, United Kingdom}

\begin{document}

\title{Testing Chern-Simons Modified Gravity with Gravitational-Wave Detections of Extreme-Mass-Ratio Binaries}

\author{Priscilla Canizares}
\email{pcm@ast.cam.ac.uk} 
\affiliation{\IAO}
\affiliation{\IEEC}

\author{Jonathan R. Gair}
\email{jrg23@cam.ac.uk}
\affiliation{\IAO}

\author{Carlos F. Sopuerta}
\email{sopuerta@ieec.uab.es} 
\affiliation{\IEEC}

\date{\today}

\preprint{}

\begin{abstract} 
The detection of gravitational waves from extreme-mass-ratio inspirals (EMRIs), comprising a stellar-mass 
compact object orbiting around a massive black hole, is one of the main targets for low-frequency 
gravitational-wave detectors in space, like the Laser Interferometer Space Antenna (LISA) or evolved
LISA/New Gravitational Observatory (eLISA/NGO). The long-duration gravitational-waveforms emitted by 
such systems encode the structure of the strong 
field region of the massive black hole, in which the inspiral occurs. The detection and analysis of 
EMRIs will therefore allow us to study the geometry of massive black holes and determine whether their 
nature is as predicted by General Relativity and even to test whether General Relativity is the 
correct theory to describe the dynamics of these systems. To achieve this, EMRI modeling in 
alternative theories of gravity is required to describe the generation of gravitational waves. However, 
up to now, only a restricted class of theories has been investigated. In this paper, we explore to 
what extent  EMRI observations with a space-based gravitational-wave observatory like LISA or eLISA/NGO might be able 
to distinguish between General Relativity and a particular modification of it, known as Dynamical 
Chern-Simons Modified Gravity.  Our analysis is based on a parameter estimation study that uses
approximate gravitational waveforms obtained via a radiative-adiabatic method.  In this framework,
the trajectory of the stellar object is modeled as a sequence of geodesics in the spacetime of the 
modified-gravity massive black hole.  The evolution between geodesics is determined by flux formulae 
based on general relativistic post-Newtonian and black hole perturbation theory computations. Once the 
trajectory of the stellar compact object has been obtained,  the waveforms are computed using the 
standard multipole formulae for gravitational radiation applied to this trajectory. Our analysis is 
restricted to a five-dimensional subspace of the EMRI configuration space, including a Chern-Simons 
parameter that controls the strength of gravitational deviations from General Relativity. We find that, 
if Dynamical Chern-Simons Modified Gravity is the correct theory, an observatory like LISA or even 
eLISA/NGO should be able to measure the Chern-Simons parameter with fractional errors below $5\%$.  
If General Relativity is the true theory, these observatories should put bounds on this parameter 
at the level $\xi^{1/4}< 10^{4}\,$km, which is four orders of magnitude better than current Solar 
System bounds.
\end{abstract}

\pacs{04.30.Db, 04.30.-w, 04.50.Kd, 95.30.Sf, 97.10.Sj}

\maketitle

\section{Introduction}
\label{intro}
There is strong observational evidence for the existence of black holes in galactic x-ray binary systems, 
seen as ultraluminous x-ray sources, and in the centers of galaxies, seen as active galactic nuclei 
(see, e.g.~\cite{Psaltis2011}). Indeed, observations carried out by space- and ground-based telescopes 
suggest the presence of  a dark compact object, likely a massive black hole (MBH), at the center of most 
observed galaxies (see~\cite{AmaroSeoane:2007aw} and references therein). 
In a typical galaxy, the MBH is surrounded by around $10^7 - 10^8$ stars forming a cusp or core (see, 
e.g.~\cite{Amaro2011}). As a consequence of relaxation, mass segregation and large scattering encounters 
between the stars, stellar compact objects (SCOs) may be perturbed onto orbits that pass sufficiently close 
to the MBH and become gravitationally bound forming a binary system. Therefore, the capture of a SCO
by a MBH is likely to be a frequent phenomenon in the Universe. 

Once the SCO has become bound to the MBH, it starts a slow inspiral driven by the emission of gravitational
waves (GWs). During this process, the system loses energy and angular momentum and the orbit of the SCO 
circularizes and shrinks adiabatically, i.e.~on a timescale much longer than the orbital period. The loss 
of energy and angular momentum occurs initially in bursts, when the object passes through the orbital 
pericenter, but eventually the gravitational radiation is being emitted continuously until the object 
reaches the innermost stable orbit and plunges into the MBH. For EMRIs whose GW frequencies lie in the 
sensitivity band of space-based GW detectors, like LISA (Laser Interferometer Space 
Antenna~\cite{Danzmann:2003tv,Prince:2003aa}) or eLISA/NGO (evolved LISA/New Gravitational 
Observatory~\cite{AmaroSeoane:2012km,AmaroSeoane:2012je}), the central MBH must have mass in the range, 
$M^{}_{\bullet} \sim 10^4-10^7 M^{}_{\odot}$. The systems of interest must also have a SCO compact enough 
to avoid tidal-disruption and so the SCO must be a stellar mass black hole ($m^{}_{\star} \approx 1-50 M_{\odot}$), 
a neutron star ($m^{}_{\star} \approx 1.4 M_{\odot}$) or a white dwarf ($m^{}_{\star} \approx 0.6 M_{\odot}$). 
The typical mass ratios, $\mu=m^{}_{\star}/M^{}_{\bullet}$, of EMRI systems are therefore in the range
$\sim 10^{-6}-10^{-4}$.

The strongest detectable EMRI signals are unlikely to be any closer than a luminosity distance 
$D \sim 1\,$Gpc~\cite{Sigurdsson:1996uz}, at which distance the instantaneous amplitude of the measured 
EMRI signal is an order of magnitude below the level of instrumental noise and the GW foreground from galactic 
white-dwarf binaries. EMRI detection will therefore rely on matched filtering of the detected data stream with a 
bank of templates of the possible signals that might be present in the data.
During the last year before plunge, an EMRI will generate $\sim 1/\mu$ gravitational waveform cycles in 
the LISA band~\cite{Finn:2000sy}. During this time, the orbit of the SCO tracks the strong field geometry 
in the vicinity of the MBH and maps out the (multipolar) structure of the MBH 
spacetime~\cite{Ryan52} in the emitted GWs.

GWs from EMRIs are generated in the strong field region close to the MBH and therefore probe General 
Relativity (GR) in a regime which, up to now, has not been reached observationally. If GR is the true 
theory of gravity describing EMRI dynamics, their waveforms will determine the parameters of the system 
with very high precision. However, if the central MBH is not described by the Kerr metric or GR does not 
properly describe the binary dynamics in the strong field regime, and we assume GR when constructing our 
detection templates, we will obtain incorrect results from GW observations. Therefore, there is a strong 
motivation for studying what kind of modifications to the dynamics of EMRIs one could expect from 
considering well-motivated theories of gravity other than GR. To that end, we must understand how the 
signals are modified in these alternative theories so that we are able to detect and quantify deviations 
from GR.

The use of EMRI observations for such tests of fundamental physics has been explored by several authors 
(see~\cite{Sopuerta:2010zy} and references therein), but the majority of that work has focussed on using 
the observations to constrain the properties of ``bumpy'' black holes.  These are solutions to the field 
equations of general relativity that represent spacetimes that differ from the Kerr solution by an amount 
controlled by a tunable deviation parameter.  EMRI observations will be able to place bounds on the size 
of deviations of the forms considered~\cite{collinshughes,glambabak,gairlimandel,BCBumpy,VigelandHughes}. 
However, this is not necessarily a test of general relativity, since the bumpy black holes are constructed 
within that theory. It is rather a test of the ``no-hair'' property of black holes (stationary astrophysical 
black holes are described by the 2-parameter (mass and spin) family of spacetime geometries of 
Kerr~\cite{Kerr:1963ud}) and hence the auxiliary assumptions that go into the no-hair conjecture. Hence, 
``bumpy'' black holes  are actually a test of the Kerr geometry assuming  GR is the correct theory of gravity.  

Due to the myriad of alternative theories of gravity available, the questions that arise are: Which kind of 
theory do we choose to compare against? What new features might we expect to observe in the GW signals 
that might allow us to distinguish this theory from GR? In this paper we address these different questions 
and explore the capability of a space-based detector like LISA to discriminate between GR and an alternative 
theory of gravity.  In particular, we focus in a modification of GR constructed by the addition of a 
Chern-Simons (CS) gravitational term (also known as the Pontryagin invariant) to the action. Interest in 
this theory was initiated with the work of Jackiw and Pi~\cite{jackiw:2003:cmo} where gravitational parity 
violation was investigated.  Such a term appears in four-dimensional compactifications of perturbative string 
theory  due to the Green-Schwarz anomaly-canceling mechanism~\cite{Polchinski:1998rr} and also in 
loop quantum gravity when the Barbero-Immirzi parameter is promoted to a scalar field coupled to the Nieh-Yan 
invariant~\cite{Taveras:2008yf,Calcagni:2009xz,Mercuri:2009zt}.  Moreover, the Pontryagin term is 
unavoidable in an effective field theory (see~\cite{Weinberg:2008hq} in the context of cosmological inflation).
In the approach of Jackiw and Pi, the Pontryagin term is introduced in the action 
multiplied by a scalar function and, in this way, it contributes to the field equations (in a four-dimensional
spacetime the Pontryagin term is a topological invariant and hence does not contribute to the field
equations), but this field is not dynamical. That is, it is a given function of the spacetimes coordinates.
This version of CS modified gravity has been extensively studied and it has been shown to be dynamically too 
restrictive and, for instance, generic oscillations of non-rotating Schwarzschild black hole are not 
allowed~\cite{Yunes:2007ss}. In addition there are problems with the uniqueness of solutions of the 
theory~\cite{Yunes2009}. For these reasons we focus on the version of the theory in which the CS scalar field 
is dynamical, i.e. Dynamical Chern-Simons Modified Gravity (DCSMG), see~\cite{Alexander:2009} for a review 
of CS modified gravity.  

The first study of EMRIs in DCSMG was done in~\cite{Sopuerta:2009iy}, where the main ingredients of the 
problem were discussed and a simple waveform model was put forward.  This model used the so-called 
{\em semi-relativistic} approximation, in which the trajectories are geodesics and the waveforms are built 
by using a standard multipolar expansion of the gravitational radiation.  Then, differences between the 
GR and DCSMG waveforms were studied and also some predictions for the relative dephasing of the waves were 
made.  However, this work relied on the assumption that radiation reaction (RR) effects, i.e.~the effects 
that arise from the interaction of the SCO with its own gravitational field would allow one to distinguish 
between GR and DCSMG.  Without RR the harmonic structure of the waveforms is going to be very similar and 
hence it is likely that it would be always possible to match a signal with both GR and DCSMG 
template waveform models.  
On the other hand, in a recent study~\cite{Pani:2011xj}, corrections to the gravitational- and scalar-wave 
fluxes for circular orbits around a non-rotating MBH in CS gravity have been computed using perturbation theory.  
This type of computations are very promising and can complement the work we present in this paper. 

In this paper we go beyond the model of~\cite{Sopuerta:2009iy} by including two 
important additional ingredients: (i) RR effects based on a hybrid scheme~\cite{Gair2006} that combines (post-Newtonian) PN 
approximations and fits to Teukolsky results~\cite{Hughes:1999bq};  (ii) Fisher parameter estimation 
techniques to make predictions on the capability of a space-based detector to measure the EMRI parameters, 
in particular a CS parameter that controls the deviations from GR. We have built {\em kludge} waveforms in 
the spirit of~\cite{Babak:2006uv} and have used them to estimate expected measurement errors for the main 
parameters describing an EMRI system in DCSMG. We find that for LISA these error estimations have the 
following order of magnitude: central black hole mass, $\Delta\log M^{}_{\bullet} \sim 5\cdot10^{-3}\,$; 
central black hole spin, $\Delta a\sim 5\cdot10^{-6}M^{}_{\bullet}\,$; orbital eccentricity, 
$\Delta e^{}_{0}\sim 3\cdot10^{-7}\,$; luminosity distance of source, $\Delta\log( D_L/\mu)\sim 2\cdot10^{-2}\,$; 
and for the CS parameter, $\xi$, in the combination $\zeta = a \xi$, we find 
$\Delta\log\,\zeta\sim 4\cdot10^{-2}\,$. Moreover, we also use this framework to put bounds on the CS 
parameter, $\xi$, directly.  Assuming that GR is the correct theory to describe ERMIs, we find that
LISA measurements could put bounds of the order $\xi^{1/4} < 10^4\,$km, which are better by four
orders of magnitude than those derived from frame dragging observations around the Earth~\cite{AliHaimoud:2011fw}.

This paper is organized as follows. In Section~\ref{formulation} we describe all the components used for 
the construction of EMRI gravitational waveforms in DCSMG and the response of space-based GW detectors. 
This includes the basic aspects of the theory, the deviations in the MBH geometry and its impact in the 
orbital dynamics and the inclusion of RR effects. In Section~\ref{signal_analysis} we summarize the 
basics elements of signal analysis theory and parameter estimation based on Fisher matrix techniques.  
In Section~\ref{Results} we apply these techniques to the waveforms and response models built in 
Sec.~\ref{formulation}, providing parameter error estimates for both LISA and eLISA/NGO and also bounds 
to the CS parameter. We finish in Section~\ref{conclusions} with conclusions and a discussion.  
Appendix~\ref{lisaspds} contains the form of the power spectral density of LISA and eLISA/NGO, while 
Appendix~\ref{evolconstantsofmotion} contains the formulae needed for the construction of the RR effects.

Throughout this paper we use Einstein summation convention for repeated indices and 
geometrized units in which $G=c=1$. Spacetime indices are denoted by Greek letters; spatial indices are 
denoted with Latin letters $i,j,\ldots$; $\nabla^{}_{\mu}$
denotes the canonical metric covariant derivative operator and $\square \equiv \met^{\mu\nu}\nabla^{}_{\mu}
\nabla^{}_{\nu}$ denotes the d'Alambertian wave operator.

\section{EMRIs in DCSMG} \label{formulation}
In order to carry out parameter estimation studies to assess the ability of a given GW detector to detect 
and extract the physical information of an EMRI system, we first need a theoretical model of the generated 
waveforms. EMRIs are complex systems and  we do not have yet a description accurate enough to produce 
waveforms in GR that can be used for data analysis purposes. However, for parameter estimation studies 
it is enough to have a waveform model that contains all the features of the real waveforms and that 
approximates the waveform phase to  within a few cycles over the whole inspiral.

Due to the large difference between the masses of the two components in an EMRI, the GW signal can be 
modeled accurately using perturbation theory (see e.g.~\cite{Drasco:2005kz}), where the SCO is represented 
as a structureless particle orbiting in the MBH spacetime background. Although on short timescales the 
orbit of the SCO is approximately a geodesic of the MBH spacetime, its parameters slowly change with 
time due to RR effects. The best method we have to estimate these RR effects is the
so-called {\em self-force} approach.  At present, the gravitational self-force has been computed 
for the case of a non-rotating MBH~\cite{Barack:2009ey,Barack:2010tm} and progress is being made towards 
calculations for the more astrophysically relevant case of a spinning MBH~\cite{Shah:2010bi} 
(see~\cite{Barack:2009ux,Poisson:2011nh,Thornburg:2011qk} for reviews).  

In parallel to the self-force program, some efforts to build certain approximation schemes to model EMRIs
have been made. For the purposes of this work we focus on the so-called {\emph{Numerical Kludge}} waveform 
model~\cite{Babak:2006uv}.  In that framework, the orbital motion is given by a sequence of geodesics around 
a Kerr MBH, with the evolution of the geodesic parameters dictated by a dissipative RR
prescription. This prescription is based on PN evolution equations for the orbital elements
(from 2PN expressions for the fluxes of energy and angular momentum) calibrated to more accurate 
Teukolsky fluxes with $45$ fitting parameters~\cite{Gair2006}. The waveforms are then modeled using a 
multipolar expansion~\cite{Thorne1980R}. 

To accurately compute the GW emission from EMRIs in an alternative theory of gravity, we need to understand 
both how the orbital dynamics of the binary are altered and how gravitational wave generation and propagation 
differs in the alternative theory. In DCSMG, the GW emission formulae are not modified at leading 
order~\cite{Sopuerta:2009iy}, and so in this paper we will consider modifications to the underlying orbital 
dynamics only.  In what follows we describe the main components of our waveform model, summarizing the 
procedure introduced in~\cite{Sopuerta:2009iy} and including the RR effects just described.

\subsection{Formulation of DCSMG}
In DCSMG the action functional depends on the spacetime metric $\met^{}_{\mu\nu}$, on the CS scalar
field $\vartheta$, and on the matter fields $\mb{\psi}^{}_{\rm mat}$, and it can be cast in the following form
\begin{eqnarray}
S[\met^{}_{\mu\nu},\vartheta,\mb{\psi}^{}_{\rm mat}] &= &\kappa^{}_{\NN}\,S^{}_{\rm EH}[\met^{}_{\mu\nu}] 
+ \frac{\alpha}{4}\,S^{}_{\rm CS}[\met^{}_{\mu\nu},\vartheta] \nonumber \\
&+& \frac{\beta}{2}\,S^{}_{\vartheta}[\met^{}_{\mu\nu},\vartheta] 
+ S^{}_{\rm mat}[\met^{}_{\mu\nu},\mb{\psi}^{}_{\rm mat}]\,,\label{action}
\end{eqnarray}
where $\kappa^{}_{\NN}$ is the gravitational constant, $1/(16\pi)$ in geometrized units, and  $\alpha$ and 
$\beta$ are universal coupling constants that control the strength of the CS modifications.
The different contributions to the action are: the GR Einstein-Hilbert action 
\begin{eqnarray}
S^{}_{\rm EH} = \int d^4x \sqrt{-\met}\,R\,, \label{EH}
\end{eqnarray}
where $\met$ is the metric determinant and $R$ is the Ricci curvature scalar; the CS gravitational
correction
\begin{eqnarray}
S^{}_{\rm CS} =  \int d^4x \sqrt{-\met}\,\vartheta\;{}^{\ast}RR\,, \label{CScorrection}
\end{eqnarray}
where ${}^{\ast}RR:= {}^{\ast}R^{\alpha}{}^{}_{\beta}{}^{\gamma\delta}R^{\beta}{}^{}_{\alpha\gamma \delta}=
\frac{1}{2}\epsilon^{\gamma\delta\mu\nu}R^{\alpha}{}^{}_{\beta\mu\nu}R^{\beta}_{\ \alpha\gamma \delta}$ is the  
Pontryagin density, $R^{\mu}{}^{}_{\nu\alpha\beta}$ is the Riemann tensor, $\epsilon^{\mu\nu\alpha\beta}$ 
is the Levi-Civita antisymmetric tensor and here the asterisk denotes the dual operation; 
the CS scalar field action term
\begin{eqnarray}
S^{}_{\vartheta} = -\int d^4x \sqrt{-\met}\,\left[ \met^{\mu \nu}
\left(\nabla_{\mu} \vartheta\right) \left(\nabla_{\nu} \vartheta\right) + 2 V(\vartheta) \right]\;, 
\label{CSaction}
\end{eqnarray}
where $V$ is the scalar field potential, which is neglected in this work (i.e. 
$V=0$); and finally, $S^{}_{\rm mat}[\met^{}_{\mu\nu},\mb{\psi}^{}_{\rm mat}]$
is the action of the different matter fields.  

Varying the action with respect to the metric and the CS scalar field, we obtain the field equations
of DCSMG:
\begin{eqnarray}
G^{}_{\mu \nu} + \frac{\alpha}{\kappa^{}_{\NN}} C^{}_{\mu \nu} & = &  
\frac{1}{2 \kappa^{}_{\NN}} \left(T_{\mu \nu}^{\rm mat} 
+ T_{\mu \nu}^{(\vartheta)}\right)\,,  \label{EEs} \\
\beta \square \vartheta &=&  - \frac{\alpha}{4} \pont\,,
\label{EOM}
\end{eqnarray}
where $G^{}_{\mu\nu}$ is the Einstein tensor and $C^{\mu \nu}$ is the so-called {\em C-tensor} which has
two  parts,  $C^{\mu \nu} = C^{\mu \nu}_{1} + C^{\mu \nu}_{2}$ with
\begin{eqnarray}
\label{Ctensor}
C^{\alpha \beta}_{1} &=& \left(\nabla^{}_{\sigma} \vartheta\right) \epsilon^{\sigma \delta \nu(\alpha}
\nabla^{}_{\nu}R^{\beta)}{}^{}_{\delta}\,, \nonumber \\
C^{\alpha \beta}_{2} &=& \left(\nabla^{}_{\sigma}\nabla^{}_{\delta}\vartheta \right) 
{\,^\ast\!}R^{\delta (\alpha \beta)\sigma}\,.
\end{eqnarray}
Finally, $T_{\mu \nu}^{\rm mat}$ is the matter stress-energy tensor and $T_{\mu \nu}^{(\vartheta)}$ is 
the stress-energy of the CS scalar field, given by 
\begin{equation}
T_{\mu \nu}^{(\vartheta)} = \beta \left[ (\nabla^{}_{\mu} \vartheta) (\nabla^{}_{\nu}\vartheta) - \frac{1}{2} 
\met_{\mu \nu}(\nabla^{\sigma} \vartheta) (\nabla^{}_{\sigma} \vartheta) \right]\,.
\label{vartheta-Tab}
\end{equation}
One can see that taking the divergence of the field equations~(\ref{EEs}), using the Bianchi identities
and the conservation of the matter stress-energy tensor, one obtains the field equation~(\ref{EOM}) for
the CS scalar field.

There are several consequences of DCSMG that are relevant for this work. The first one is that the number 
of independent waveform polarizations that a detector far away from a GW source will see are the same 
in DCSMG as in GR~\cite{Sopuerta:2009iy}, i.e., the plus and cross tensor GW polarizations. In addition 
to the plus and cross polarizations, in DCSMG there is an additional breathing mode, however it decays 
faster, typically like $r^{-2}$ and is therefore unlikely to be detected by an observer far away from 
the source. Another important property of GWs in DCSMG is the structure of the stress-energy (or mass) 
tensor that can be associated with the GWs in the {\em short-wave} approximation 
(see, e.g.~\cite{Misner:1973cw}), commonly known as the Isaacson 
tensor~\cite{Isaacson:1968ra,Isaacson:1968gw}.  In~\cite{Sopuerta:2009iy} it was shown that the GW 
stress-energy tensor has the same form (in terms of the gauge-invariant metric perturbation describing
the GWs) as the one of Isaacson for GR. This is due to the fact that the averaging involved in the 
short-wave approximation cancels out all the CS corrections giving rise, at leading order, to 
essentially the same backreaction in DCSMG as in GR.

\subsection{The MBH geometry in DCSMG}
The first ingredient we need to model the dynamics of an EMRI system is the geometry of the MBH.  In GR 
we know that, provided the {\em no-hair} conjecture is true, all MBHs must be described by the Kerr 
metric.  However, this is no longer true in DCSMG.  We do not have an exact solution in DCSMG for 
spinning MBHs, but there is an approximate solution~\cite{Yunes2009,Konno:2009kg} that has been found 
using a small-coupling approximation (using 
$\zeta^{}_{\rm CS}\equiv \alpha^{2}/(M^{}_{\bullet}\beta\kappa^{}_{\NN})$ as the expansion parameter, 
with $M^{}_{\bullet}$ being the MBH mass) and a slow-rotation approximation (defined by 
$a/M^{}_{\bullet}\ll 1$, with $a \equiv |\mb{S^{}_{\bullet}}|/M^{}_{\bullet}$, 
$0\leq a/M^{}_{\bullet}\leq 1$, and $\mb{S}^{}_{\bullet}$ is MBH spin). 
Using a system of coordinates that in the GR limit coincide with the well-known Boyer-Lindquist (BL) 
coordinates $(t,r,\theta,\phi)$~\cite{Yunes2009}, the non-vanishing metric components have the 
following form
\begin{eqnarray}
{\met}^{}_{tt} & = & -\left(1-\frac{2M^{}_{\bullet}r}{\rho^2}\right)\,, \label{gtt} \\
{\met}^{}_{rr} & = & \frac{\rho^2}{\Delta}\,,\\
{\met}^{}_{\theta \theta} & = & \rho^2 \,,\\
{\met}^{}_{\phi \phi} & = & \frac{\Sigma}{\rho^2}\sin^2{\theta}\,,\\
{\met}^{}_{t\phi} & = &\left[\ \frac{5}{8} \frac{\xi}{M_{\bullet}^{4} } \frac{a}{M^{}_{\bullet}}
\frac{M^5_{\bullet}}{r^4}\left(1+\frac{12M_{\bullet}}{7r} + \frac{27M^2_{\bullet}}{10r^{2}}\right)\right.
\nonumber \\
&&\left.- \frac{2M^{}_{\bullet}a r}{\rho^2}\ \right]\sin^2\theta\;\,,\label{DCSMG_metric}
\end{eqnarray}
where we have introduced the following definitions: $\rho^2=r^2+a^2\cos^2\theta$, 
$\Delta = r^2f+a^2$, $f = 1-2M^{}_{\bullet}/r$, and $\Sigma=(r^2+a^2)^2-a^2\Delta\sin^2\theta$.
The effects of the CS gravitational modification are parametrized by a single universal constant,
$\xi$, given by
\begin{eqnarray}
\xi:=\frac{\alpha^2}{\beta\kappa^{}_{\NN}}\,.\label{xi}
\end{eqnarray}
Notice that the only metric component that gets modified with respect to the general relativistic 
case is the component ${\met}^{}_{t\phi}$ [Eq~(\ref{DCSMG_metric})].  The term in this component 
that is proportional to the CS parameter $\xi$ falls off with distance as $r^{-4}$, that is, it 
decays much faster than the rest of the metric components and hence it becomes negligible at 
large distances. Only gravitational systems like EMRIs can probe this modification as they 
penetrate into the strong field region of the MBH.

At the level of approximation at which the DCSMG metric [Eq~(\ref{DCSMG_metric})] was obtained, 
it is possible to show that it has most of the properties of the Kerr metric~\cite{Sopuerta:2009iy}, 
in particular the DCSMG metric is stationary and axisymmetric, and also has a Killing tensor, 
which is important for an analysis of the orbital motion. Moreover,  the DCSM metric has the same 
algebraic structure as the Kerr one.  Regarding the multipolar structure of this DCSMG metric, 
let us remember that the multipole moments of the Kerr metric are fully determined by the MBH 
mass and spin (or equivalently, by the mass monopole and current dipole) according to the 
following simple relations: $M_{\ell} + i S_{\ell} = M \left(i a\right)^{\ell}$, where 
$\{M_{\ell}\}^{}_{\ell=0,\ldots,\infty}$ and $\{S_{\ell}\}^{}_{\ell=0,\ldots,\infty}$ are 
the mass and current multipole moments respectively.  The multipole moments associated with the 
CS metric deviate from those of Kerr starting at the  $S^{}_{4}$ multipole, 
as one can see by employing the multipolar formalism of~\cite{Thorne1980R} (see also~\cite{Gursel}).
Despite this deviation, the structure of these multipole moments still preserves the philosophy
of the no-hair conjecture since they only depend on the mass and spin of the MBH.  There is also
a dependence on the CS parameter $\xi$, but this is a universal dependence that would be the same
for all MBHs and hence it cannot be considered to be {\em hair} of the MBH.

The equations for the metric and CS scalar field are coupled, so they have to be solved simultaneously.
The solution for the CS scalar, at the same level of approximation as for the metric, is:
\begin{equation}
\vartheta =  \frac{5}{8} \frac{\alpha}{\beta} \frac{a}{M^{}_{\bullet}} \frac{\cos\theta}{r^2} 
\left(1 + \frac{2 M^{}_{\bullet}}{r} + \frac{18 M^2_{\bullet}}{5 r^2} \right)\,.
\label{back-theta}
\end{equation}
This scalar field falls off as $r^{-2}$ and therefore it has a finite energy associated with it.

\subsection{Orbital Kinematics}\label{kinematics}
It was argued in~\cite{Sopuerta:2009iy} that in DCSMG massive particles should follow geodesics of
the spacetime metric.  At the lowest order of approximation, and for short periods of time, the
trajectory of the SCO can then be approximated by geodesics of the metric given in 
Eqs.~(\ref{gtt})-(\ref{DCSMG_metric}). Actually, we are going to approximate the orbital motion 
as a sequence of geodesics, as in the GR case within the Numerical Kludge (NK) waveform model.  For this reason, it 
is important to analyze in detail the structure of the geodesic motion around this 
modified-Kerr metric.

In the previous subsection we mentioned that the modified MBH geometry has essentially the
same physical and geometrical properties as the Kerr metric.  In particular it has the same
number of symmetries.  Therefore, we can separate the geodesic equations
as in the Kerr case, introducing certain constants of the motion.  More specifically,
we have the energy per unit SCO  mass, $E$, the angular momentum component along the spin axis
per unit SCO mass, $L^{}_{z}$, and finally the Carter constant per unit SCO mass squared, $C$.

The geodesic equations have the following structure~\cite{Sopuerta:2009iy}:
\begin{eqnarray}
\dot{t} & = & \dot{t}^{}_{\rm K} + L_z\,\delta g^{\rm CS}_{\phi}(r)\,,\label{geoCS_1} \\
\dot{\phi} & = &\dot{\phi}^{}_{\rm K} - E\,\delta g^{\rm CS}_{\phi}(r)\,,\label{geoCS_1b}\\
\dot{r}^2 & = & \dot{r}^2_{\rm K} + 2EL_zf\,\delta g^{\rm CS}_{\phi}(r)\,, \label{geoCS_2}\\ 
\dot{\theta}^2 & = & \dot{\theta}^2_{\rm K}\,, \label{dottheta}
\end{eqnarray}
where the dots denote differentiation with respect to proper time.  The quantities 
($\dot{t}^{}_{\rm K}$, $\dot{r}^{}_{\rm K}$, $\dot{\theta}^{}_{\rm K}$, 
$\dot{\phi}^{}_{\rm K}$) are the counterparts of the
geodesic equations in the Kerr metric, which are given by (see, e.g.~\cite{Chandrasekhar:1992bo})
\begin{eqnarray}
\rho^2\dot{t}^{}_{\rm K} &=& -a\left(a\,E\sin^2\theta-L^{}_z\right)\nonumber \\
&+&\frac{r^2+a^2}{\Delta}\left[\left(r^2+a^2\right)E-aL^{}_z \right]\,,\label{geoK_1} \\
\rho^2\dot{\phi}^{}_{\rm K} &=&  \frac{a}{\Delta}\left[(r^2+a^2)E-aL^{}_z\right]
-\left(aE-\frac{L^{}_z}{\sin^2\theta}\right) \,, \label{geoK_1b} \\
\rho^4\dot{r}_{\rm K}^2 &=& \left[(r^2 + a^2)E -aL_z\right]^2 \nonumber\\
&-&\Delta\left[ Q +\label{geoK_2}(aE - L_z)^2 +r^2\right],\\
\rho^4\dot{\theta}^2_{\rm K} &=&  Q - \cot^2\theta L_z^2 - a^2\cos^2\theta(1 - E^2)\,,
\label{geoK_3}
\end{eqnarray}
where $Q$ is an alternative definition of the Carter constant, related to $C$ by
\begin{equation}
Q = C -  \left(L^{}_{z} - a E \right)^{2}\,. \label{relationCtoQ}
\end{equation}

It is clear from Eqs.~(\ref{geoCS_1})-(\ref{geoCS_2}) that the CS deviations are determined 
by a single function of the radial coordinate $r$, $\delta g^{\rm CS}_{\phi}(r)$, which has 
the form 
\begin{eqnarray}
\delta g^{\rm CS}_{\phi}= \frac{\xi a}{112r^6\,f}\left(70+120\frac{M^{}_{\bullet}}{r}
+189\frac{M^2_{\bullet}}{r^{2}} \right)\,.\label{cs_correction}
\end{eqnarray}
Only the equation for the polar coordinate $\theta$ is unchanged.
Since we are dealing with bound orbits, both the radial and the polar motion
include turning points (extrema of motion) at which the time derivatives, $\dot{r}$ or 
$\dot{\theta}$, vanish. This can create numerical problems when integrating the set of 
Ordinary Differential Equations (ODEs) given by Eqs.~(\ref{geoCS_1})-(\ref{geoCS_2}).  
To avoid this, we follow the same strategy as in the case of Kerr geodesics and
introduce two angle coordinates, $\psi$ and $\chi$, associated with the radial and polar 
motion respectively:
\begin{eqnarray}
r = \frac{pM^{}_{\bullet}}{1+e\cos\psi}\,,~~~~~\cos^2\theta = \cos^2\theta_{\MIN}\cos^2\chi\,, 
\label{new_coordinates}
\end{eqnarray}
where $p$ and $e$ are the dimensionless semilatus rectum and the eccentricity of the orbit
respectively, and $\theta^{}_{\MIN}$ is the minimum of $\theta$ in the orbit (the turning 
point in the polar motion). The orbital parameters $(e,p)$ are related with the radial 
turning points, the apocenter ($r^{}_{\APO}$) and pericenter ($r^{}_{\PERI}$), through 
the standard expressions:
\begin{eqnarray}
r^{}_{\PERI} = \frac{pM^{}_{\bullet}}{1+e}\,,\quad
r^{}_{\APO} = \frac{pM^{}_{\bullet}}{1-e}\,.
\end{eqnarray}
or equivalently
\begin{equation}
p = \frac{2\,r^{}_{\PERI}\,r^{}_{\APO}}{M^{}_{\bullet}(r^{}_{\PERI}+r^{}_{\APO})}\,, \qquad
e = \frac{r^{}_{\APO}-r^{}_{\PERI}}{r^{}_{\PERI}+r^{}_{\APO}}\,.
\end{equation}
The radial coordinate $r$ oscillates in the interval $(r^{}_{\PERI},r^{}_{\APO})$.
Similarly, given the turning point of the polar motion, $\theta^{}_{\MIN} \in [0,\pi/2]$, 
$\theta$ performs a libration motion in the interval $(\theta^{}_{\MIN},\pi-\theta^{}_{\MIN})$). 
We introduce the orbital inclination angle (with respect to the spin direction), through the 
following relation
\begin{equation}
\theta^{}_{\INC} = {\rm sign}(L^{}_{z})\left[\frac{\pi}{2} - \theta^{}_{\MIN}\right]\,, 
\label{theta-inc}
\end{equation}
where ${\rm sign}(L^{}_{z}) =1$ corresponds to a prograde orbit and ${\rm sign}(L^{}_{z}) =-1$ 
corresponds to a retrograde orbit.  A different definition of the orbital inclination angle 
can be given in terms of the constants of motion $(E,L^{}_{z},C$ or $Q)$
\begin{equation}
\cos\iota = \frac{L^{}_{z}}{\sqrt{L^{2}_{z}+Q}}\,.  \label{iota-angle}
\end{equation}
In general, both inclination angles, $\theta^{}_{\INC}$ and $\iota$, are quite 
similar~\cite{Drasco:2003ky} and coincide in the non-spinning limit, $a=0$.

We work with two geodesic parameterizations, one based on the orbital parameters 
$(e,p,\theta^{}_{\INC}$ or $\iota)$ and one based on the {\em constants} of motion 
$(E,L^{}_{z},C$ or $Q)$. Changing from one parameterization to the other is a fundamental step 
in our computations. In the GR case there is a well-known procedure (see, 
e.g.~\cite{Schmidt:2002qk,Drasco:2003ky}) to do so.  Here, we have used the implementation 
described in the appendices of~\cite{Sopuerta:2011te}. However, these formulae are only valid 
in GR and, in our case, the CS modification of the radial equation of motion changes the 
location of the turning points.  In practice this translates into a different relation between 
the two sets of parameters $(e,p,\theta^{}_{\INC}$ or $\iota)$ and $(E,L^{}_{z},C$ or $Q)$.   
Given that we are not dealing with large deviations from the GR case, we have used a numerical 
procedure based on the Newton-Raphson method for finding roots (see, e.g.~\cite{Press:1992nr}), 
where the values obtained from the GR method have been used as the starting point for the 
iteration algorithm.  We have seen that in practice this works quite well and the iteration 
converges rapidly to the correct values.

Finally, due to the separability of the geodesic equations, which is closely related to
the spacetime symmetries, we can distinguish in the motion three {\em fundamental} frequencies 
(here with respect to coordinate time $t$) associated with the radial motion, 
$f^{}_{r} = 1/T^{}_{r}$ ($T^{}_{r}$ is the average time to go from pericenter to apocenter and 
back to pericenter), with the polar motion, $f^{}_{\theta} = 1/T^{}_{\theta}$ ($T^{}_{\theta}$ 
is the average time for a full oscillation of the orbital plane, going from 
$\theta=\theta^{}_{\MIN}$ to $\theta=\pi-\theta^{}_{\MIN}$ and back to $\theta=\theta^{}_{\MIN}$), 
and $f^{}_{\phi}=1/T^{}_{\phi}$ ($T^{}_{\phi}$ is the average time for the SCO's azimuthal 
angular coordinate $\phi$ to cover $2\pi$ radians).  It is important to mention that these 
frequencies change when including the CS modifications ~\cite{Sopuerta:2009iy}.

\subsection{Orbital Dynamics}\label{radreact}
So far, the orbital dynamics described  have been for geodesic orbits. In order to compute the 
SCO trajectory, we gradually evolve the parameters of the instantaneous geodesic orbit under RR. 
The RR effects not only drive the SCO inspiral, but also break the degeneracy between orbits 
in GR and others in DCSMG~\footnote{Without RR effects the phase evolution of an EMRI waveform, 
in both theories, will be a multiple Fourier series of the three fundamental frequencies, 
i.e.~it will contain harmonics of the type $\exp\{2\pi if^{}_{m,n,p}t\}$ with 
$f^{}_{m,n,p} = m f^{}_{r} +n f^{}_{\theta}+ p f^{}_{\phi}$.  Then, we would be able to 
associate physical parameters with a given EMRI in both theories, and hence we would not be 
able to discriminate between them.}, since a given initial orbital configuration will evolve 
differently in these theories.  Therefore we need to implement RR effects in the EMRI dynamics 
in the framework of DCSMG.

As we have mentioned above, in this paper we adapt the NK waveform model to the case of 
DCSMG~\cite{Babak:2006uv}. In the NK waveform model, the RR driven evolution uses a ``hybrid'' 
scheme described in~\cite{Gair2006}, where formulae for the evolution of the {\em constants} of 
motion $(E,L^{}_{z},C$ or $Q)$ are derived in terms of PN approximations (at 2PN order) combined 
with fits to results from the Teukolsky formalism (see~\cite{Hughes:1999bq,Hughes:2001jr}).
In principle one should then derive the analogous formulae for the case of DCSMG, but this 
involves a number of major developments that are currently out of reach.  Instead, we will take 
into account one of the important results about GWs in DCSMG discussed previously --- the 
realization that the stress-energy momentum tensor for GWs, the Isaacson tensor, has the same 
form in terms of the GW metric perturbation in both theories, GR and DCSMG.  This means that, 
to leading-order, the properties of the GW emission in GR and DCSMG are the same.  Then, we 
approximate the fluxes of energy and angular momentum in the GWs, and also the evolution of 
the Carter constant under GW emission, by using the GR expressions.  In what follows, we 
describe the formulae and procedures to update the geodesic orbits in our NK-EMRI model.

The evolution equations for the constants of motion $(E,L^{}_{z},C$ or $Q)$ have the
following structure:
\begin{eqnarray}
\frac{{\rm d}E}{{\rm d}t} &=& \mu \,f^{}_E(a,p,e,\theta^{}_{\INC})\,,\label{edot_eq} \\
\frac{{\rm d}L^{}_z}{{\rm d}t} &=& \mu \, f^{}_{L^{}_z}(a,p,e,\theta^{}_{\INC})\,, \label{lzdot_eq} \\
\frac{{\rm d}Q}{{\rm d}t} &=& \mu \, f^{}_Q(a,p,e,\theta^{}_{\INC}) \,. \label{qdot_eq}
\end{eqnarray}
The evolution equation for $C$ follows from these equations and Eq.~(\ref{relationCtoQ}).   
The form of the right-hand sides $f^{}_{E}$, $f^{}_{L^{}_{z}}$, and $f^{}_{Q}$ of Eqs. (~\ref{edot_eq})-(\ref{qdot_eq}) 
and full details of their derivation can be found in~\cite{Gair2006}.  In Appendix~\ref{evolconstantsofmotion} 
we summarize the main expressions needed to build these right-hand sides and thus evaluate the evolution
of the constants of motion.

In practice, there are two ways in which we can use the evolution equations~(\ref{edot_eq})-(\ref{qdot_eq}). 
The first one consists of computing a phase-space trajectory for the orbital parameters 
by integrating the set of ODEs for the evolution of the energy, $E$, 
angular momentum component along the spin axis, $L^{}_z$, and Carter constant, $Q$.  Once the 
phase space trajectory ($E(t)$, $L^{}_z(t)$, $Q(t)$) has been computed, these time-dependent constants 
are used on the right-hand side of the geodesic equations~(\ref{geoK_1})-(\ref{geoK_3}), to construct 
the inspiral trajectory of the SCO in the Boyer-Lindquist-like coordinates of the MBH spacetime.
The second option, the one that we use in this paper, is to consider the extended system of
ODEs consisting of Eqs.~(\ref{geoK_1})-(\ref{geoK_3}) and Eqs.~(\ref{edot_eq})-(\ref{qdot_eq})
and integrate them together in time.  Although this system of ODEs is coupled, there is a clear
hierarchical structure, since the subsystem of Eqs.~(\ref{edot_eq})-(\ref{qdot_eq}) can in principle
be integrated independently of the subsystem of Eqs.~(\ref{geoK_1})-(\ref{geoK_3}), which can be
seen as a subsidiary system.

As mentioned before, Eqs.~(\ref{edot_eq})-(\ref{qdot_eq}) are in principle only valid in GR.
If the true theory of gravity is DCSMG, these evolution equations will contain corrections. 
At leading order, GW emission in DCSMG takes the same form as in GR~\cite{Sopuerta:2009iy}, but 
corrections to the fluxes will still arise from the DCSMG modifications to the orbital motion. 
These corrections were computed for circular orbits in DCSMG in~\cite{GairYunes11}, but enter at a 
high post-Newtonian order. For this reason, we do not make any modifications to the GR expressions 
but directly employ the fluxes described in Gair and Glampedakis~\cite{Gair2006}. Although we therefore 
use the same RR formulae to evolve the trajectory in DCSMG as in GR, this still leads 
to a different SCO evolution, since the dependence of the orbital elements $(e,p,\theta^{}_{\INC}$ or $\iota)$
on the ``constants'' of motion $(E,L_z,C$ or $Q)$ is different in the two theories, which leads to 
correspondingly different gravitational waveforms.  That is,
the mapping between the orbital elements $(e,p,\theta^{}_{\INC}$ or $\iota)$ and the {\em constants} of
motion $(E,L^{}_{z},C$ or $Q)$ is different in DCSMG and GR.  Therefore, given some initial orbital 
parameters $(e^{}_{0},p^{}_{0},\theta^{}_{\INC,0}$ or $\iota^{}_{0})$, after evolving the EMRI system for some time, 
the final orbital parameters in GR will be in general different from the orbital parameters in DCSMG.  

\begin{figure}[htb]
\centering
\includegraphics[width=0.49\textwidth]{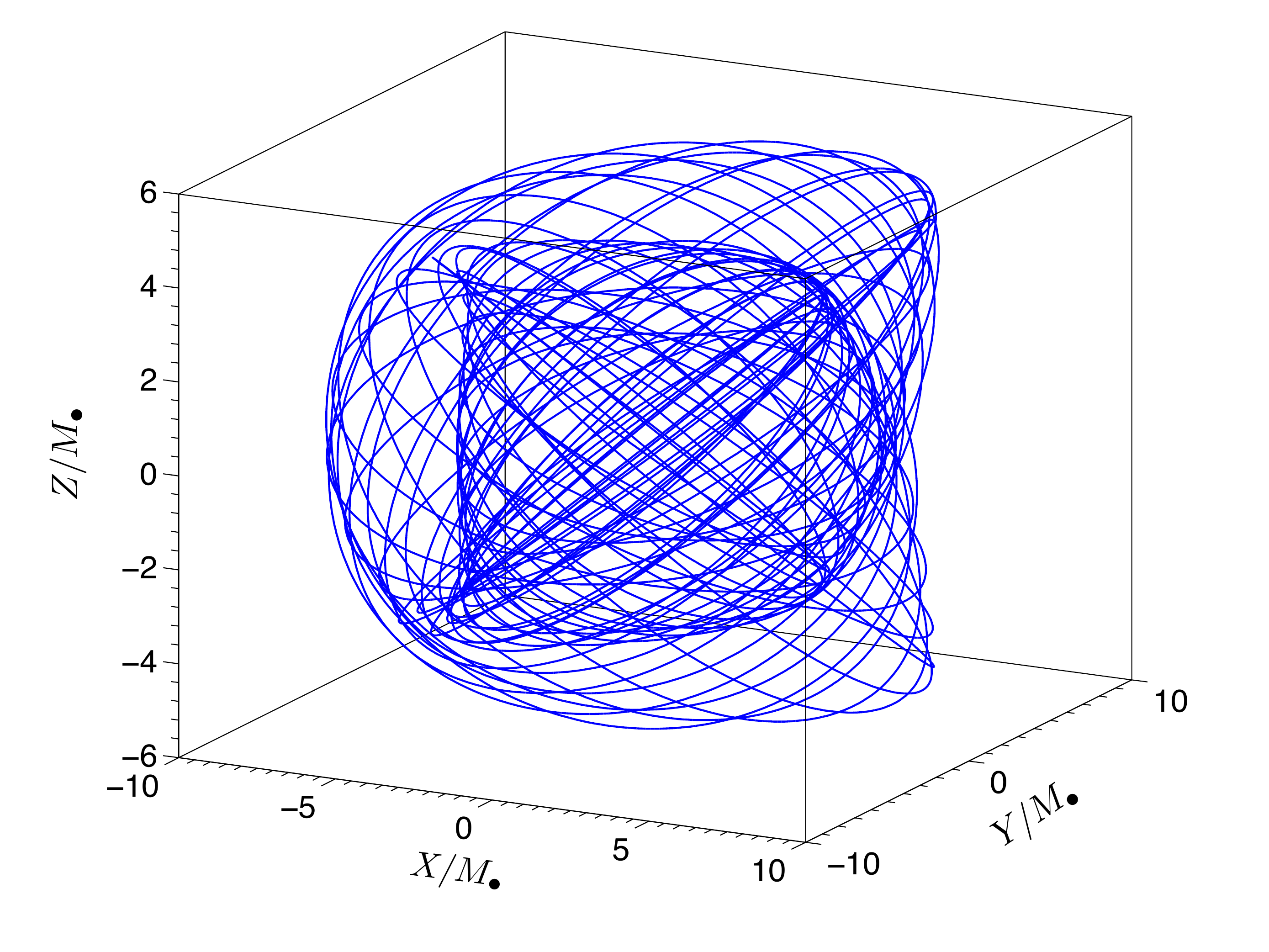}
\caption{Depiction of a section of the inspiral orbit for an EMRI (system $A$ in Table
~\ref{EMRI_systems}) with parameters   $M^{}_{\bullet} = 5\cdot10^5M^{}_{\odot}$, $a = 0.25M^{}_{\bullet}$,
$e_0 = 0.25$ and $\zeta = 5\cdot10^{-2}M^{5}_{\bullet}$.
\label{orbitex}}
\end{figure}

Taking into account the previous considerations, the inspiral is constructed in the following way: 
For a given set of initial orbital parameters $(e^{}_{0},p^{}_{0},\theta^{}_{\INC,0}$ or $\iota^{}_{0})$, 
we find the associated initial constants of the motion 
$(E^{}_{0},L^{}_{z,0},C^{}_{0}$ or $Q^{}_{0})$, which are different from the ones that we would obtain in 
GR for the same initial eccentricity, semilatus rectum and inclination angle. Subsequently, we evolve 
the ``constants'' of motion, $(\dot{E}|^{}_{0}, \dot{L}^{}_z|^{}_{0},\dot{Q}|^{}_{0})$, under RR, 
using the method described above. Then, from the current values of the constant 
of motion, $(E^{}_{0},L^{}_{z,0},C^{}_{0}$ or $Q^{}_{0})$, their rates of change due to RR, 
$(\dot{E}|^{}_{0}, \dot{L}^{}_z|^{}_{0},\dot{Q}|^{}_{0})$, and the value of the radial period, $T_r$ 
(the time to go from the apocenter to the pericenter and back again to apocenter)~\cite{Sopuerta:2009iy},  
we obtain the new constants of motion, $(E^{}_{1},L^{}_{z,1},Q^{}_{1})$, using the following equations:
\begin{eqnarray}
E^{}_{1} & = & E^{}_{0} + \dot{E}|^{}_{0}\,N^{}_r\,T^{}_r\,,\\
L^{}_{z,1} & = & L^{}_{z,0} + \dot{L}_{z}|^{}_{0}\,N^{}_r\, T^{}_r\,,\\
Q^{}_{1} & = & Q^{}_{0} + \dot{Q}|^{}_{0}\,N^{}_r\,T^{}_r\,,
\end{eqnarray}
and $N^{}_r$ is a pre-specified parameter that represents the number of radial periods elapsed 
between each update of the constants of the motion.  The expression for $C^{}_{1}$ follows from 
these formulae for $(E^{}_{1},L^{}_{z,1},Q^{}_{1})$ and Eq.~(\ref{relationCtoQ}).
Finally from  $(E^{}_{1},L^{}_{z,1},C^{}_{1}/Q^{}_{1})$, 
we obtain the new values of the orbital parameters $(e^{}_{1},p^{}_{1},\theta^{}_{\INC,1}/\iota^{}_{1})$. 
This algorithm is iterated along the whole EMRI evolution to obtain the SCO orbit.  
In Figure~\ref{orbitex}
we illustrate a section of a generic orbit for a typical system that we use later in our parameter 
estimation analysis (see Table~\ref{EMRI_systems}).

\subsection{Waveform Modeling and Detector Responses}\label{waveform-and-responses}
In the previous subsections we have seen how the trajectory of the SCO is obtained and, in the 
following, we describe how we compute the gravitational waveforms and the response of the LISA and 
eLISA/NGO detectors. Following~\cite{Babak:2006uv} and~\cite{Sopuerta:2009iy} we employ the 
multipolar expansion of the metric perturbations describing the GWs emitted by an isolated system, 
which assumes that the GWs propagate in a flat background spacetime to reach the 
observer/detector~\cite{Thorne1980R}. In this work, we consider only the lowest-order term, the 
mass quadrupole.  This term involves second time derivatives of the trajectory and these are readily 
obtained from the geodesic equations~(\ref{geoK_1})-(\ref{geoK_3}).  Then, the transverse-traceless 
(TT) metric perturbation is computed from the following expression 
\begin{eqnarray}
h^{\TT}_{ij}(t) = \frac{2}{r}  \ddot{I}^{}_{ij}\,,\label{multipolar_hij}
\end{eqnarray}
where $I^{}_{ij}$ denotes the mass quadrupole and $r$ the luminosity distance from the source 
to the observer. In terms of the source stress-energy tensor, $T^{}_{\mu\nu}$, the mass quadrupole 
moment is:
\begin{eqnarray}
I^{ij} &=& \left[\int d^{3}x\; x^i\, x^j\; T^{tt}(t,x^{i})\right]^{{\rm STF}}, \label{massquadrupole}
\end{eqnarray}
where ${\rm STF}$ stands for symmetric and trace-free.   Treating the SCO in the point-mass approximation,
the non-vanishing components of the stress energy tensor have the following form: $T^{tt}(t,x^{i}) =\rho(t,x^{i})$ 
and $T^{tj}(t,x^{i}) = \rho(t,x^{i}) v^{j}(t)$, where $\rho(t,x^{i})$ is the energy density of the SCO
which, in the point-mass limit, is given by:
\begin{eqnarray}
\rho(t,x^i) = m^{}_{\star}\, \delta^{(3)}\left[x^i-z^i(t)\right]\,,\label{rhoxt}
\end{eqnarray}
where $\delta^{(3)}$ denotes the three-dimensional Dirac delta distribution, $z^i(t)$ are the spatial Cartesian 
coordinates (associated with the flat spacetime
background) of the SCO trajectory and $v^i(t)=dz^i(t)/dt$ are the components of the corresponding
spatial velocity.
To evaluate this in our model we make a ``particle-on-a-string'' 
approximation, that is we identify the Boyer-Lindquist-like coordinates $(r,\theta,\phi)$ of the SCO's orbit 
with flat-space spherical polar coordinates, and introduce Cartesian coordinates in the usual way
\begin{eqnarray}
x = r\sin\theta\cos\phi\,,\quad
y = r\sin\theta\sin\phi\,,\quad
z = r\cos\theta\,.
\end{eqnarray}
Although this description leads to inconsistencies, like the non-conservation
of the flat-space energy-momentum tensor of
the particle motion, it has been found to work well when generating EMRI waveforms in GR~\cite{Babak:2006uv} 
and so we do not expect this to introduce large errors in the waveforms, 
in particular in the phase.   A possible alternative could be to use coordinate systems more adapted to
the multipolar expansion of the gravitational radiation, like harmonic or asymptotic-Cartesian mass-centered 
coordinates~\cite{Thorne1980R} (see also~\cite{Sopuerta:2011te}).

We now consider detection of these signals by a space-based detector in a heliocentric orbit (like LISA or eLISA/NGO).  
We describe the direction from the detector to the EMRI system by a unit 3-vector $\hat{\textbf{n}}$, which also gives
the propagation direction of the GWs from the EMRI to the detector.  The orthogonal plane to $\hat{\textbf{n}}$
is the GW polarization plane and we can introduce there two unit and orthogonal vectors $\hat{\textbf{p}}$ and 
$\hat{\textbf{q}}$ by using the spin direction, $\mb{S}^{}_{\bullet} = a M^{}_{\bullet}\hat{\textbf{z}}$:
\begin{eqnarray}
\hat{\textbf{p}}= \frac{\hat{\textbf{n}}\times\hat{\textbf{z}}}{|\hat{\textbf{n}}\times\hat{\textbf{z}}|}\,,\qquad
\hat{\textbf{q}}=\hat{\textbf{p}}\times\hat{\textbf{n}}\,.
\end{eqnarray}
The vectors $(\hat{\textbf{n}},\hat{\textbf{p}},\hat{\textbf{q}})$ form a spatial orthonormal basis that 
can be used to construct the GW polarization tensors:
\begin{eqnarray}
\epsilon_{+}^{ij}= p^{}_ip^{}_j -q^{}_iq^{}_j\,,\qquad
\epsilon_{\times}^{ij} = 2p^{}_{(i}q^{}_{j)}\,.
\end{eqnarray}
The corresponding \emph{plus}, $h^{}_{+}$, and \emph{cross}, $h^{}_{\times}$, GW polarizations are 
given by:
\begin{eqnarray}
h^{}_{+}(t) =\frac{1}{2}\epsilon_{+}^{ij}h^{}_{ij}(t)\,, \qquad
h^{}_{\times}(t)=\frac{1}{2}\epsilon_{\times}^{ij}h^{}_{ij}(t)\,,
\end{eqnarray}
and the complete GW metric perturbation is: 
\begin{eqnarray}
h^{}_{ij}(t)=\epsilon^{+}_{ij}\,h^{}_{+}(t) + \epsilon^{\times}_{ij}\,h^{}_{\times}(t)\,.
\end{eqnarray}
Using Eqs.~(\ref{multipolar_hij}) and~(\ref{rhoxt}) we obtain the following simplified expressions 
for the GW polarizations in terms of the SCO position $z^{i}(t)$, velocity $v^{i}(t) = dz^{i}/dt$,
and acceleration $a^{i}(t) = d^{2}z^{i}/dt^{2}$:
\begin{eqnarray}
h^{}_{+,\times}(t) = \frac{2 m^{}_{\star}}{r} \epsilon_{ij}^{+,\times} \left[ a^{i}(t) z^{j}(t)
+ v^{i}(t) v^{j}(t)\right]\,.
\end{eqnarray}

Once we have the GW waveforms, we compute the response function of a space-based detector in heliocentric motion.
Due to the motion of the LISA and eLISA/NGO constellations as they orbit (rotation and translation) 
it is more convenient to rewrite the response functions in terms 
of angles defined in a fixed Solar System Barycenter (SSB) coordinate system.  
The direction from the origin of the SSB reference frame 
to the origin of the EMRI reference frame is $-\hat{\textbf{n}}$
\begin{eqnarray}
-\hat{\textbf{n}} =-(\sin\theta^{}_{\rm S}\cos\phi^{}_{\rm S}, \sin\theta^{}_{\rm S}\sin\phi^{}_{\rm S},
\cos\theta^{}_{\rm S})\,,
\end{eqnarray}
where $(\theta^{}_{\rm S},\phi^{}_{\rm S})$ are spherical polar angles that determine the sky location
of the EMRI with respect to the SSB frame.  The relations between these angles and the angles
$(\theta^{}(t),\phi^{}(t))$ that determine the sky location with respect to the detector
reference frame are (see, e.g.~\cite{Cutler:1998cc,Barack:2003fp}):
\begin{eqnarray}
\cos\theta^{}(t) &=& \frac{1}{2}\cos\theta^{}_{\rm S}
-\frac{\sqrt{3}}{2}\sin\theta^{}_{\rm S}\cos(2\pi t/T-\phi^{}_{\rm S}) \,, \label{LISAtheta} \\[1mm]
\phi^{}(t)& =& 2\pi t/T +\Phi(t)\,,
\end{eqnarray}
where
\begin{eqnarray}
\Phi(t) = \tan^{-1}\left[\frac{\sqrt{3}\cos\theta^{}_{\rm S}+\sin\theta^{}_{\rm S}\cos\left(2\pi t/T 
-\phi^{}_{\rm S}\right)}{2\sin\theta^{}_{\rm S}\sin\left(2\pi t/T -\phi^{}_{\rm S}\right)}\right]\,,
\end{eqnarray}
and $T=1$yr is the period of the Earth's orbit around the Sun.  The polarization angle $\psi$ describes the orientation of the ``apparent 
ellipse" given by the projection of the orbit on the sky.  It can be written in terms of 
$(\theta^{}_{\rm S},\phi^{}_{\rm S})$  and the angles describing the direction of the MBH spin with respect 
to the SSB reference frame $(\theta^{}_{\rm K},\phi^{}_{\rm K})$:
\begin{eqnarray}
\tan \psi &=& \left[ \left\{ \cos\theta^{}_{\rm K} -
\sqrt{3}\sin\theta^{}_{\rm K}\cos\left(2\pi t/T -\phi^{}_{\rm K}\right)\right\} \right.\nonumber\\ \nonumber
&&\left. -2\cos\theta(t)\left\{\cos\theta^{}_{\rm K}\cos\theta^{}_{\rm S} \right.\right.\\ \nonumber
&&\left.\left. +\sin\theta^{}_{\rm K}\sin\theta^{}_{\rm S}\cos(\phi^{}_{\rm K}-\phi^{}_{\rm S}) \right\} \frac{}{}\right]/\\ \nonumber
&&  \left[\frac{}{} \sin\theta^{}_{\rm K}\sin\theta^{}_{\rm S}\sin(\phi^{}_{\rm K}-\phi^{}_{\rm S})\right.\\ \nonumber
&&\left. -\sqrt{3}\cos\left(2\pi t/T \right)\left\{\cos\theta^{}_{\rm K}\sin\theta^{}_{\rm S}\sin\phi^{}_{\rm S}\right.\right.\\ \nonumber
&&\left.\left. -\cos\theta^{}_{\rm S}\sin\theta^{}_{\rm K}\sin\phi^{}_{\rm K} \right\} \right.\\  \nonumber
&&\left. -\sqrt{3}\sin\left(2\pi t/T) \left\{\cos\theta^{}_{\rm S}\sin\theta^{}_{\rm K}\cos\phi^{}_{\rm K}
\right.\right.\right.\\
&&\left.\left. -\cos\theta^{}_{\rm K}\sin\theta^{}_{\rm S}\cos\phi^{}_{\rm S} \right\} \frac{}{}\right]\,.\label{LISApsi}
\end{eqnarray}
In addition, the time of arrival of a gravitational wavefront at the SSB and at the detector will 
in general differ and are related by
\begin{eqnarray}
t^{}_{\rm SSB} = t^{}_{\rm D} + R\,\sin\theta^{}_{\rm S} \cos(2\pi t^{}_{\rm D}/T -\phi^{}_{\rm S}) -
t_{\rm SSB}^0 \,,
\end{eqnarray}
where $R = 1\,$ AU, $t^{}_{\rm D}$ is the time of arrival as seen in the detector reference frame, 
and $t_{\rm SSB}^0$ is the initial time in the SSB reference frame:
\begin{eqnarray}
t_{\rm SSB}^0 = t^0_{\rm D} + R\,\sin\theta^{}_{\rm S} \cos(2\pi t^0_{\rm D}/T -\phi^{}_{\rm S}) \,.
\end{eqnarray}
This difference in arrival times gives rise to a Doppler modulation in the GW phase measured 
by LISA. To compute a waveform regularly sampled in time at the detector, we need to generate 
a waveform unevenly sampled in the source frame (in which the time sampling is the same as at 
the SSB). This can be achieved employing the relations just introduced.

The response of the detector to an incident GW can then be written as:
\begin{eqnarray}
h^{}_{\alpha}(t)=\frac{\sqrt{3}}{2}\left[F^{+}_{\alpha}(t)h^{}_{+}(t)+F^{\times}_{\alpha}(t)h^{}_{\times}(t)\right]\,,
\end{eqnarray}
where $\alpha$ is an index for the different independent channels of the detector.
In the case of LISA we have two independent Michelson-like interferometer channels 
that can be constructed from the LISA data stream and hence $\alpha=I,II$.  By contrast,
eLISA/NGO will have only one independent channel (see Appendix~\ref{lisaspds} for a brief comparison
of the detectors) and hence $\alpha=I$ for eLISA/NGO.  The antenna pattern (response) functions, 
$F_{\alpha}^{+}{}^{,\times}$, are given by (see, e.g.~\cite{Barack:2003fp}):
\begin{eqnarray}
F^{+}_I &=& \frac{1}{2}(1+\cos^2\theta)\cos(2\phi)\cos(2\psi) \nonumber \\
&&-\cos\theta\sin(2\phi)\sin(2\psi)\,,  \\
F^{\times}_I &=& \frac{1}{2}(1+\cos^2\theta)\cos(2\phi)\cos(2\psi)\nonumber \\
&&+\cos\theta\sin(2\phi)\sin(2\psi)\,, \\
F^{+}_{II} &=& \frac{1}{2}(1+\cos^2\theta)\sin(2\phi)\cos(2\psi) \nonumber 
\\&&+\cos\theta\cos(2\phi)\sin(2\psi)\,, \\
F^{\times}_{II} &=& \frac{1}{2}(1+\cos^2\theta)\sin(2\phi)\sin(2\psi) \nonumber \\
&&-\cos\theta\cos(2\phi)\cos(2\psi)\,.
\end{eqnarray}
Here $(\theta,\phi, \psi)$ are as defined in Eqs.~(\ref{LISAtheta})--(\ref{LISApsi})
\mbox{ }
and specify the sky location and orientation of the source in a detector-based coordinate system 
in terms of angles defined in a fixed SSB coordinate system.

\section{Elements of Signal Analysis and Model Parameter Estimation}\label{signal_analysis}
The starting point for signal analysis is the detector data stream(s), $s^{}_{\alpha}\,$.  We assume that 
$s^{}_{\alpha}$ contains an EMRI GW signal, $h^{}_{\alpha}$, and hence we can decompose it as
\begin{equation}
s^{}_\alpha(t) = h^{}_\alpha(t) + n^{}_\alpha(t)\,,
\end{equation}
where $n^{}_\alpha(t)$ is the noise in the detector, which we assume to be stationary, Gaussian 
and, in the case of LISA, that the two data streams are uncorrelated and the noise power spectral
density is the same in each channel.  Then, the Fourier components of the noise, which we denote
with a tilde $\tilde{n}^{}_\alpha(f)$ (see~\cite{Cutler:1998cc,Barack:2003fp} for conventions 
on the Fourier transform that we use), satisfy
\begin{eqnarray}
\langle \tilde{n}^{}_\alpha (f)\tilde{n}^{\ast}_\beta (f')\rangle = \frac{1}{2} 
\delta^{}_{\alpha\beta} \delta(f-f') S^{}_n(f) \,,
\end{eqnarray}
where $\langle\cdot\rangle$ denotes expectation value (ensemble average over all possible realizations 
of the noise), the asterisk now denotes complex conjugation, and $S^{}_{n}(f)$ is the (one-sided) power 
spectral density (PSD) of  the noise, which is given in Appendix~\ref{lisaspds} for both LISA and 
eLISA/NGO. The assumption of Gaussian noise means that the probability of
a particular realization of the noise ${\bf n}^{}_0$ is given by
\begin{eqnarray}
p({\bf n}={\bf n}^{}_0) \propto e^{-({\bf n}^{}_0|{\bf n}^{}_0)/2}\,, \label{prob}
\end{eqnarray}
where $(\cdot |\cdot)$ denotes
the natural inner product in the vector space of signals associated with the PSD $S^{}_{n}(f)$
and is defined as
\begin{eqnarray}
(\textbf{a}|\textbf{b}) = 2\sum^{}_{\alpha} \int_0^\infty {\rm d}f\,
\frac{\tilde{a}^{\ast}_\alpha(f) \tilde{b}^{}_\alpha(f) + \tilde{a}^{}_\alpha(f)
\tilde{b}^{\ast}_\alpha(f)}{S^{}_n(f)} \,, \label{s-overlap}
\end{eqnarray}
for any two signals $\textbf{a}$ and $\textbf{b}$.  The probability that a given GW signal 
$\textbf{h}$ is present in a data stream $\textbf{s}$ is  thus
\begin{equation}
p(\textbf{s}|\textbf{h})\propto e^{-(\textbf{s}-\textbf{h}|\textbf{s}-\textbf{h})/2}\,.
\label{likelihood2}
\end{equation}
The `best-fit' waveform will be the one that 
maximizes $(\textbf{s}|\textbf{h})$ and, thus, it provides the maximum likelihood parameter 
estimate. The expected signal-to-noise ratio (SNR), when filtering with the correct waveform, is
\begin{eqnarray}
\text{SNR}=\frac{(\textbf{h}|\textbf{h})}{{\rm rms}(\textbf{h}|\textbf{n})} = \sqrt{(\textbf{h}|\textbf{h})}\,, 
\label{SNR}
\end{eqnarray}
where `rms' stands for {\em root mean square} and the second inequality follows from the fact that the 
expectation value of $(\textbf{a}|\textbf{n})(\textbf{b}|\textbf{n})$ is 
$(\textbf{a}|\textbf{b})$~\cite{Cutler:1994ys}.   In practice one considers a waveform template
family that will depend on a set of parameters $\mb{\lambda}$, $\{\textbf{h}(t,\mb{\lambda})\}$,
and searches for the parameters that maximize the probability of a certain noise realization, i.e., the
probability that a given waveform template is present in the data stream.   
Different realizations of the noise will lead to different values of the best-fit parameters.  
For large SNR, the best-fit parameters will follow a Gaussian distribution centered around
the correct values.  Expanding  $\exp(-(\textbf{s}-\textbf{h}|\textbf{s}-\textbf{h})/2)$ around the 
best-fit parameters,  $\mb{\lambda}^{}_0$, by writing $\mb{\lambda} =  \mb{\lambda}^{}_0 + 
\delta \mb{\lambda}$, we obtain the following form
for the probability distribution function for the errors $\delta\mb{\lambda}$ 
\begin{eqnarray}
p(\delta\mb{\lambda}) = {\cal N} \exp\left(-\frac{1}{2} \Gamma^{}_{jk} 
\delta\lambda^j\delta\lambda^k\right)\,,\label{likelihood}
\end{eqnarray}
where ${\cal N}=\sqrt{{\rm det} ( \Gamma /2 \pi)}$ is the normalization factor and $\Gamma_{ij}$ 
is the \emph{Fisher information matrix} (FM)~\cite{Fisher:1935}
\begin{eqnarray}
\Gamma^{}_{jk} =\left( \frac{\partial \textbf{h}}{\partial \lambda^j}\, \vline\,\frac{\partial 
\textbf{h}}{\partial \lambda^k} \right)\,.\label{FM}
\end{eqnarray}
The variance-covariance matrix for the waveform parameters is given by the inverse of the FM:
\begin{eqnarray}
\langle \delta\lambda^j\delta\lambda^k \rangle = \left( \Gamma^{-1}\right)^{jk} \left[1
+ O(1/{\rm SNR})\right]\,,\label{covariance}
\end{eqnarray}
and hence, we can estimate the precision with which we will be able to measure a particular 
parameter, $\lambda^{i}$, by computing the component $\Gamma^{-1}_{ii}$ of this inverse matrix,
that is (see~\cite{Vallis2008} for a detailed discussion):
\begin{equation}
\Delta\lambda^{i}\equiv \sqrt{\left\langle \left(\delta\lambda^{i}\right)^{2}\right\rangle} 
\simeq \sqrt{\Gamma^{-1}_{ii}}\,. \label{estimated-error}
\end{equation}
%

\subsection{The Maximum-Mismatch Criterion}\label{MMC}
Vallisneri~\cite{Vallis2008} provided a consistency criterion to determine whether the SNR is 
high enough for the FM results to be trustworthy, called the \emph{Maximum-Mismatch Criterion} 
(MMC). The MMC criterion was suggested to assess when an estimation of the parameter errors based 
on a FM analysis would be reliable or not. Since the FM, $\Gamma^{}_{ij}$, is built from the 
partial derivatives of the waveform template with respect to the parameters of the model, it can 
only represent the true GW signal, $h^{}_{\rm GW}$, correctly if $h(t,\mb{\lambda})$ is linear 
in all the parameters, $\mb{\lambda}$, across a parameter space region of size comparable to the 
expected parameter errors. This is the regime in which the \emph{Linearized-Signal Approximation} 
(LSA) is valid. As we increase the SNR the errors become smaller and consequently the LSA is 
expected to work better.  In the regime of validity of the LSA we can expand the 
waveform template $h(t,\mb{\lambda})$ around the {\em true} source parameters, $\mb{\lambda}^{}_{\rm tr}$,
i.e. $\lambda^i =\lambda^i_{\rm tr}+\delta\lambda^i$ with $\delta\lambda^i$ being a small deviation 
in the parameters comparable with the parameter estimation error:
\begin{eqnarray}
h(t,\mb{\lambda}) = h^{}_{\rm tr} + \delta\lambda^i\left(\partial_ih^{}\right)|^{}_{\lambda^{k}_{\rm tr}}
+\frac{\delta\lambda^i\delta\lambda^j}{2}\left(\partial^2_{ij}h\right)|^{}_{\lambda^k_{\rm tr}}
+ \ldots  \label{LSA}
\end{eqnarray}
Then, the likelihood [Eq.~(\ref{likelihood2})] can be approximated as:
\begin{eqnarray}
p(\textbf{s}|\mb{\lambda}) &\propto&
\exp\left\{-\frac{({\bf n}|{\bf n})}{2} +\delta\lambda^i\delta\lambda^j
\frac{\left(\partial^{}_i\textbf{h}|\partial^{}_j\textbf{h}\right)}{2} \right. \nonumber \\
& + & \left. \delta\lambda^j \left(\partial^{}_j\textbf{h}|\bf{n}\right) \frac{}{}\right\}\,,
\label{likelihood_LSA}
\end{eqnarray}
where the waveform template derivatives are evaluated at $\mb{\lambda}=\mb{\lambda}^{}_{\rm tr}$.
The applicability of the FM for parameter estimation is limited by the high-SNR requirement, 
in the sense that it can be a poor predictor of the amount of information obtained from 
waveforms depending on several parameters and detected with relatively low SNR. The  MMC is given in terms of the
ratio, $r$, of the LSA likelihood [Eq.~(\ref{likelihood_LSA})] to the exact likelihood 
[Eq.~(\ref{prob})]:
\begin{eqnarray}
2|\log r| = \left(\Delta\lambda^i(\partial^{}_i\textbf{h})^{}_{\lambda^k_{\rm tr}} - 
\textbf{Dh} \left| \Delta\lambda^j
(\partial^{}_j \textbf{h})^{}_{\lambda^k_{\rm tr}} - \textbf{Dh}\right)\,, \right.
\label{logr}
\end{eqnarray}
where $\textbf{Dh} = \textbf{h}(\lambda^k_{\rm tr}+\Delta\lambda^k) - \textbf{h}(\lambda^k_{\rm tr})$ 
and $\Delta\lambda^i$  is the estimated error from the diagonal components of the inverse of the FM. 
The MMC is obtained by taking the maximum value of $r$ over all parameters.

The idea behind the MMC is to choose an iso-probability surface as predicted by the FM, 
and explore it to verify that the difference between the LSA and exact likelihoods is 
sufficiently small. Ratios, $r$, below some fiducial value are considered acceptable. 
If this condition is satisfied, we can believe that the FM is providing a reliable 
estimate of the parameter estimation errors.

\section{Parameter Estimation Studies: Methods and Results}\label{Results}
In this section we describe the different techniques employed in our parameter estimation 
analysis and present the main results.  We begin by characterizing the EMRI parameter
space in our studies in DCSMG, $\{\lambda^i\}^{}_{i=1,\ldots,N}$.  There are  
15 parameters (the 14 of GR plus the DCSMG coupling parameter; see Table~\ref{parameters} 
for a brief description): 
$\mb{\lambda} = \{M^{}_{\bullet}, a, \mu, e^{}_0, p^{}_0, \theta^{}_{\INC,0}, \zeta, 
\theta^{}_{\rm S}, \phi^{}_{\rm S}, \theta^{}_{\rm K}, \phi^{}_{\rm K} ,D^{}_{\rm L}, 
\psi^{}_0, \chi^{}_0, \phi^{}_0\}$, where the subscript $0$ refers to the values of the 
corresponding quantities at the inspiral initial time. 

\begin{table}
\caption{Summary of the parameters that characterize an EMRI system in DCSMG. The angles 
$(\theta^{}_{\rm S},\phi^{}_{\rm S})$ and $(\theta^{}_{\rm K}, \phi^{}_{\rm K})$ are spherical 
polar coordinates with respect to the ecliptic and the subindex $0$ stands for values of 
parameters computed at the initial time. The parameters 
with physical dimensions are indicated in square brackets. We set the luminosity distance to
$D^{}_{\rm L} = 1\,$Gpc.  \label{parameters}} 
\begin{ruledtabular}
\begin{tabular}{ll}
Parameter                                    & ~Description \\
\hline\\[-2mm]
$M^{}_{\bullet}$                             & ~MBH mass [$M_{\odot}$]. \\
$a = |\mb{S}^{}_{\bullet}|/M^{}_{\bullet}$~  & ~MBH Spin [$M^{}_{\bullet}$]. \\
$\mu = m^{}_{\star}/M^{}_{\bullet} $ 	        & ~EMRI mass ratio. \\
$e^{}_0$                                     & ~Eccentricity of the particle orbit at $t^{}_0$.\\
$p^{}_0$                                     & ~Dimensionless semilatus rectum at $t^{}_0$.\\
$\theta^{}_{\INC,0}$                         & ~Inclination of the orbit at $t^{}_0$.\\
$\zeta$                                      & ~$\xi\cdot a$~~[$M_{\bullet}^{5}$].\\
$\theta^{}_{\rm S}$                          & ~EMRI polar angle.\\
$\phi^{}_{\rm S}$                            & ~EMRI azimuthal angle.\\
$\theta^{}_{\rm K}$                          & ~MBH spin polar angle.\\
$\phi^{}_{\rm K}$                            & ~MBH spin azimuthal angle.\\
$D^{}_{\rm L}$                               & ~Distance from the SSB to the EMRI [Gpc].\\
$\psi^{}_0$                                  & ~Angle variable for the radial motion.\\
$\chi^{}_0$                                  & ~Angle variable for the polar motion.\\
$\phi^{}_0$                                  & ~Boyer-Lindquist azimuthal angle.\\
\end{tabular} 
\end{ruledtabular}
\end{table} 

In order to simplify the computations involved in this study, we have restricted ourselves
to a five-dimensional subset of the parameter space, given by 
$\mb{\lambda} = \{M^{}_{\bullet}, a, e^{}_0, \zeta, D^{}_{\rm L}/\mu\}\,$ 
(see Table~\ref{parameters} for their definition). In this subset we have included those 
parameters that we have found to have the greatest correlation with the parameter 
$\zeta = a\cdot\xi$ (see Table~\ref{parameters}), which controls the strength of the CS 
modifications (notice that in the MBH metric of Eqs.~(\ref{gtt})-(\ref{DCSMG_metric}) the 
CS parameter $\xi$ always appears multiplied by the spin parameter $a$ and this has motivated 
the introduction of the combined parameter $\zeta$) in a full parameter space investigation. 
We have also checked that the results we obtain do not change significantly when more 
parameters are added to the FM study. 
For the parameter estimation studies we consider two different EMRI systems, $A$ and $B$,
whose parameters are given in Table~\ref{EMRI_systems}.  These two types of systems
differ in the values for the MBH mass, $M^{}_{\bullet}=5\cdot10^{5}M^{}_{\odot}$ for system
$A$ and $M^{}_{\bullet}=10^{6}M^{}_{\odot}$ for system $B$.
We fix the luminosity distance to $D^{}_{\rm L} = 1\,$Gpc,  which roughly corresponds to 
the distance where we might expect the closest 
detectable sources to lie (see, e.g.~\cite{Sigurdsson:1997vc}). Due to the fact that the 
inspiral time scales as $\sim\mu$ with the mass ratio, the system $A$ evolves faster than  
system $B$, which allows us to use smaller evolution times to obtain reliable results in 
that case.  

\begin{table}
\caption{EMRI systems considered in the parameter estimation analysis.  The table shows the values for
the parameters that are considered in the FM computation (see Table~\ref{parameters} for the whole
list of parameters).  The rest of EMRI parameters employed in our parameter estimation 
analysis are the same for both systems and their values are: $m^{}_{\star}=10M_{\odot}$, 
$\theta^{}_{\INC,0}=0.569$, $\theta^{}_{\rm S}=\phi^{}_{\rm S}=1.57$, 
$\theta^{}_{\rm K}=\phi^{}_{\rm K}=0.78$, $\psi^{}_0=\chi^{}_0=\phi^{}_0=0.78$.
\label{EMRI_systems}}  
\begin{ruledtabular}
\begin{tabular}{l|ccccc}
System~      & $M^{}_{\bullet}$  & $a/M^{}_{\bullet}$     & $e^{}_0$   & $\zeta/M^{5}_{\bullet}$  & $D^{}_{\rm L}/\mu$ [Gpc]  \\[1mm]
\hline\\[-1mm]
$A$          & $5\cdot 10^5$     & $0.25$  & $0.25$     & $5\cdot 10^{-2}$	   & $5\cdot10^{4}$\\[2mm]
$B$          & $10^6$            & $0.25$  & $0.25$     & $5\cdot 10^{-2}$	   & $10^{5}$\\
\end{tabular}
\end{ruledtabular}
\end{table} 

For these systems we evolve the trajectory using the geodesic equations given in Sec.~\ref{kinematics}  
and the RR equations given in Sec.~\ref{radreact}.  This is done using the 
algorithm outlined in Sec.~\ref{radreact}.  The ODEs that describe geodesic
motion are integrated for the angle variables $[\psi(t),\chi(t),\phi(t)]$ using
the Bulirsch-Stoer extrapolation method~\cite{Bulirsch:1966bs} (see~\cite{Press:1992nr,Stoer:1993sb} 
for details). The numerical code also contains routines which convert back and forth between 
the different parameterizations of the orbit in DCSMG;
that compute the Cartesian orbital coordinates, velocities, and accelerations, the multiple moments; 
etc. The equations that evolve the constants of motion, $E$, $L_z$ and $Q$ are integrated using  simple finite
difference rules. Then, we use the formulae of Sec.~\ref{waveform-and-responses} to compute 
the gravitational waveforms and the detector responses.

\begin{figure}[htb]
\centering
\includegraphics[width=0.5\textwidth]{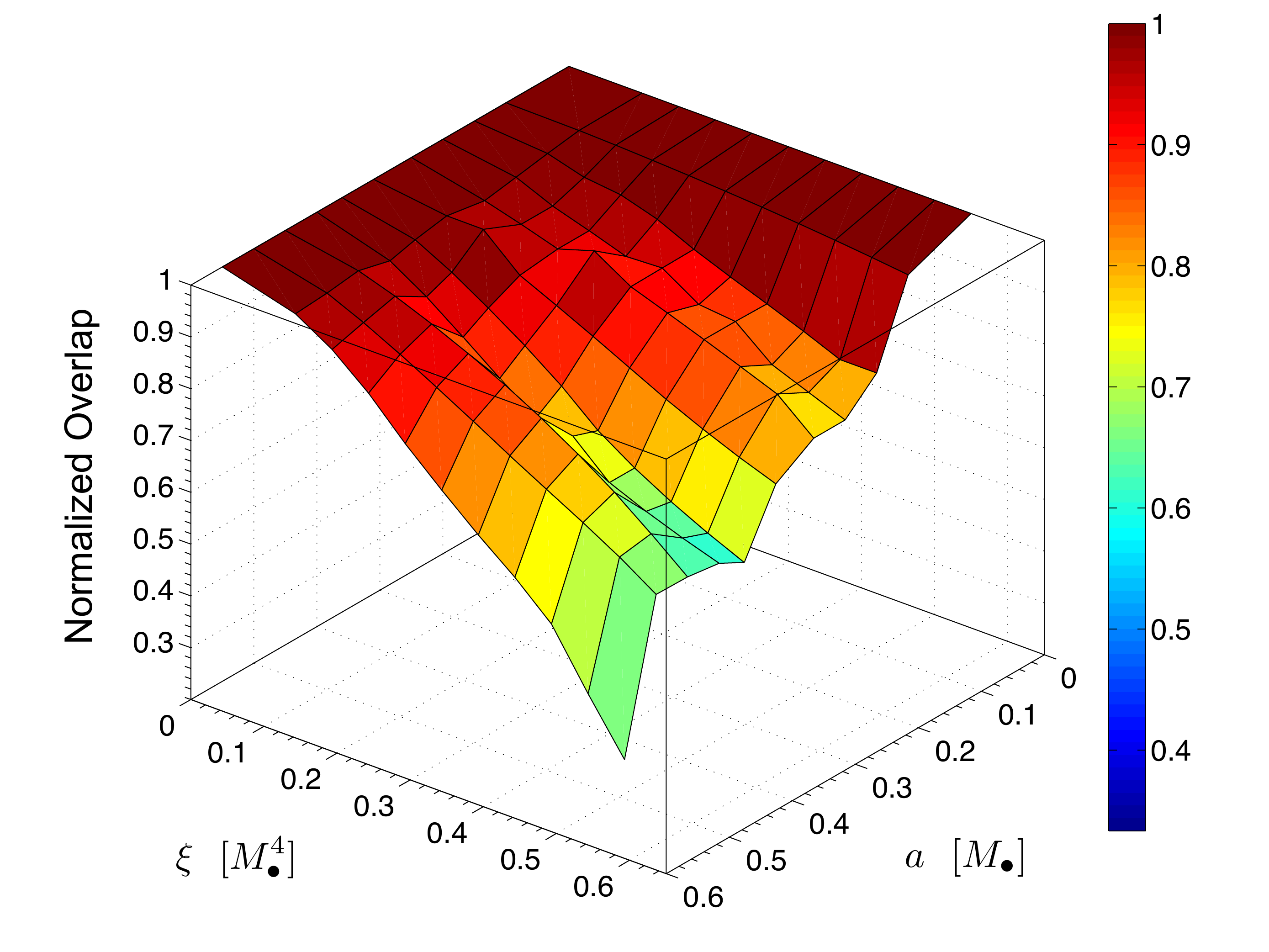}
\caption{This 2D plot shows the symmetric normalized overlap of Eq.~(\ref{o-gr-cs}) for EMRI
systems with the following parameters: $M^{}_{\bullet}=5\cdot 10^{5}\,M^{}_{\odot}\,$,
$m^{}_{\star}=10\,M^{}_{\odot}\,$, $e^{}_{0} = 0.25\,$, $p^{}_{0} = 11\,$,
$\theta^{}_{\INC,0}=0.569\,$, $\theta^{}_{\rm S} = 1.57\,$, $\phi^{}_{\rm S} = 1.57\,$,
$\theta^{}_{\rm K}=0.329\,$, $\phi^{}_{\rm K}=0.78\,$, $D^{}_{\rm L}/\mu = 5\cdot 10^{4}\,$ Gpc,
$\psi^{}_{0} = 0.78\,$, $\chi^{}_{0}=0.78\,$, $\phi^{}_{0} = 0.78\,$.  The parameters
$\xi/M^{4}_{\bullet}$ and $a/M^{}_{\bullet}$ take values in the interval $[0,0.5]$ with
a step of $0.05\,$ for a total of $121$ points. \label{overlap-GR-CS}
}
\end{figure}

In order to study how different the waveforms in DCSMG are from GR, we have evolved our EMRI 
system during $0.5$yr employing different values of the CS parameter $\xi$ and the MBH spin $a$, 
and we have computed the following overlap function between a DCSMG and a GR waveform template:
\begin{eqnarray}
{\cal O}\left[\textbf{h}^{}_{\rm GR},\textbf{h}^{}_{\rm CS}\right]\equiv
\frac{\left(\textbf{h}^{}_{\rm GR}|\textbf{h}^{}_{\rm CS} \right)}
{\sqrt{\left(\textbf{h}^{}_{\rm GR}|\textbf{h}^{}_{\rm GR} \right)
\left(\textbf{h}^{}_{\rm CS}|\textbf{h}^{}_{\rm CS} \right)}}\,, \label{o-gr-cs}
\end{eqnarray}
which is symmetric, ${\cal O}\left[\textbf{h}^{}_{\rm GR},\textbf{h}^{}_{\rm CS}\right]
= {\cal O}\left[\textbf{h}^{}_{\rm CS},\textbf{h}^{}_{\rm GR}\right]$ and has
the obvious property: ${\cal O}\left[\textbf{h}^{}_{\rm GR},\textbf{h}^{}_{\rm GR}\right]
= {\cal O}\left[\textbf{h}^{}_{\rm CS},\textbf{h}^{}_{\rm CS}\right] = 1\,$.  
We also assume that the two waveforms used for this overlap correspond to EMRIs
with the same parameters, except for the CS parameter $\zeta$ that vanishes
for GR waveforms.  The standard overlap defined in Eq.~(\ref{s-overlap}) has been computed
using the FFTW library~\cite{fftw:2005} for the Fourier transforms and simple integration rules.
We have computed this normalized overlap for a total of $121$ EMRI systems which have
13 fixed parameters: $ M^{}_{\bullet}=5\cdot 10^{5}\,M^{}_{\odot}\,$,
$m^{}_{\star}=10\,M^{}_{\odot}\,$, $e^{}_{0} = 0.25\,$, $p^{}_{0} = 11\,$,
$\theta^{}_{\INC,0}=0.569\,$, $\theta^{}_{\rm S} = 1.57\,$, $\phi^{}_{\rm S} = 1.57\,$,
$\theta^{}_{\rm K}=0.329\,$, $\phi^{}_{\rm K}=0.78\,$, $D^{}_{\rm L}/\mu = 5\cdot 10^{4}\,$ Gpc,
$\psi^{}_{0} = 0.78\,$, $\chi^{}_{0}=0.78\,$, $\phi^{}_{0} = 0.78\,$, while the spin $a/M^{}_{\bullet}$ 
and the CS parameter $\xi/M^{4}_{\bullet}$ are varied in the interval $[0,0.5]\,$.  The results 
are shown in Figure~\ref{overlap-GR-CS}, where we can see how the projection of 
$\textbf{h}^{}_{\rm CS}$ onto $\textbf{h}^{}_{\rm GR}$
changes by modifying the values of the MBH spin $a/M^{}_{\bullet}$ and the CS parameter 
$\xi/M^{4}_{\bullet}$. In particular, for higher values of $a/M^{}_{\bullet}$ and 
$\xi/M^{4}_{\bullet}$ the overlap ${\cal O}\left[\textbf{h}^{}_{\rm GR},\textbf{h}^{}_{\rm CS}\right]$ 
decreases, since the difference in the evolution of the SCO in GR and CS, produced by the  
dephasing introduced by the RR, increases [see Eqs.~(\ref{geoCS_1})-(\ref{geoCS_2})] and, 
consequently, the deviations of $\textbf{h}^{}_{\rm CS}$ from $\textbf{h}^{}_{\rm GR}$ 
are enhanced.

We have also obtained the SNR in the frequency domain using
Eq.~(\ref{SNR}).  The computation of the FM requires the evaluation of the derivatives of the
waveform templates, $\partial^{}_i\textbf{h} = \partial\textbf{h}/\partial\lambda^{i}\,$
(actually of the response functions of the detector).  
Since the waveform templates/responses are generated numerically, the corresponding
derivatives must also be evaluated numerically.  For inner points in the EMRI parameter space
(i.e. not near boundaries so that we do not need points outside the proper domains of
definition of the parameters) we use the following five-point finite-difference rule:
\begin{eqnarray}
\partial^{}_i \textbf{h} & = &  \frac{1}{12\,\delta\lambda^i}
\left\{ \textbf{h}(\lambda^i+2\delta\lambda^i) 
-\textbf{h}(\lambda^i-2\delta\lambda^i) \right. \nonumber \\ 
&+&\left. 8\, \left[ \textbf{h}(\lambda^i+ \delta\lambda^i)  - \textbf{h}
(\lambda^i-\delta\lambda^i)\right] \right\} + O\left[(\delta\lambda^{i})^{4}\right]\,, 
\label{waveformderivative}
\end{eqnarray}
where $\delta\lambda^{i}$ is the numerical offset in the parameter $\lambda^{i}$.   For 
computations near the boundary or at the boundary of the parameter space we use instead
non-centered finite-differences rules.  Either the following three-point rule
\begin{eqnarray}
\partial^{}_{i} \textbf{h} & = & \frac{1}{2\,\delta\lambda^{i}}
\left\{ 4\textbf{h}(\lambda^{i}+\delta\lambda^{i})- \textbf{h}(\lambda^{i} + 2\delta\lambda^{i}) - 
3\textbf{h}(\lambda^{i}) \right\} \nonumber \\
& + & O\left[(\delta\lambda^{i})^{2}\right]\,,~~ \label{threepointrule}
\end{eqnarray}
or the following four-point rule
\begin{eqnarray}
\partial^{}_{i} \textbf{h} & = & \frac{1}{4\,\delta\lambda^{i}} 
\left\{  \textbf{h}(\lambda^{i}+2\delta\lambda^{i}) 
- \textbf{h}(\lambda^{i}+3\delta \lambda^{i}) \right. \nonumber \\ 
&+& \left. 5\, \left[ \textbf{h}(\lambda^{i} + \delta\lambda^{i})- 
\textbf{h}(\lambda^{i})\right] \right\}+O\left[(\delta\lambda^{i})^{3}\right]\,. \label{fourpointrule}
\end{eqnarray}
It is known that computing numerical derivatives is a delicate task (see, e.g.~\cite{Press:1992nr}).  
In the case of finite difference formulas like Eq.~(\ref{waveformderivative}) the choice of the 
offset $\delta\lambda^{i}$ is crucial.  An offset too small will produce high-order cancellations 
in the numerator beyond machine precision.  In contrast, an offset too big may mean higher order 
terms in the Taylor series expansion of the waveform become important.  In both cases we will be 
far from a reasonable approximation.  Therefore, we have done investigations that survey wide 
ranges for $\delta\lambda^{i}$ in order to find intervals where the derivatives have good 
convergence properties.

Once we have obtained a FM, $\Gamma^{}_{ij}$, that converges in a certain range of offsets 
$\delta\lambda^{i}$, we estimate the expected measurement error in the parameters by using 
Eq.~(\ref{estimated-error}). Since FMs for EMRI waveforms have very large condition numbers 
(the ratio of the largest to the smallest eigenvalues), we use an \emph{LU decomposition} to
invert them, writing the matrix as the product of a lower triangular matrix and an upper 
triangular matrix~\cite{Huerta:2008gb}. In addition, to assess whether the error estimates 
obtained are reliable or not, we use the MMC defined in Eq.~(\ref{logr}).
We have evaluated the MMC criterion for all the results presented in this paper and, unless 
otherwise specified, they satisfy this criterion with values of $|\log r|$ ranging from 
$10^{-4}$ to $0.5$.

We have stated before that RR effects 
change the relative phase between waveforms in DCSMG with respect to GR.
Now, we are going to study their impact on parameter estimation. Firstly, we will compare 
the parameter estimation errors for systems 
evolved under RR and systems that do not radiate, preserving the constants of
motion, i.e. always have the same orbital parameters. These results have been obtained assuming that the
detector is LISA.  At the end of this section we present some results for eLISA/NGO. The 
results for system $A$ with and without RR and for different evolution times,  
$T^{}_{\rm evol} = 0.1\,$, $0.3\,$, $0.5\,$, $1$ yr, are shown in Table~\ref{table_RR}.  
The upper part of the table 
contains the results for the evolutions with RR and the lower part of the table shows
the results without RR.  In both cases we also show the value of the MMC test, i.e.
the quantity $|\log r|$ defined in Eq.~(\ref{logr}). 
We do not show results for $T^{}_{\rm evol}=0.1\,$yr without RR in  
Table~\ref{table_RR} since we did not obtain reliable results (according to the MMC criterion). 
From these results and others we have obtained for other similar EMRI systems
we can say that the typical measurement accuracies for the five most important parameters are: 
$\Delta\log M^{}_{\bullet}\sim 10^{-3}\,$, 
$\Delta a \sim 10^{-6}\,M^{}_{\bullet}\,$, 
$\Delta e^{}_{0}\sim 10^{-7}\,$, 
$\Delta\log\,\zeta \sim 10^{-2}$ and 
$\Delta\log (D^{}_{\rm L}/\mu)\sim 10^{-2}\,$.
Comparing the results in Table~\ref{table_RR}, we see that the inclusion 
of the RR improves the SNR of the signals.  It also improves the parameter estimates, 
in particular those of the spin, $a$, and of the CS parameter, $\zeta$. 
This is partially due to the increase of the overall SNR due to RR, but even after rescaling 
to a fixed reference SNR we see an improvement 
in the parameter measurement accuracies when RR is included. As one could expect 
due to the adiabatic nature of the RR (see e.g~\cite{Sago:2005gd}), the improvement 
with the inclusion of RR is more significant for longer evolution times.

\begin{table*}[ht]
\caption{Error estimates for LISA and the EMRI system $A$ (see Table~\ref{EMRI_systems}) using RR (upper part of the table) 
and without using RR (lower part of the table).  Each column contains the estimations for a given 
evolution time ($T^{}_{\rm evol} = 0.1\,$, $0.3\,$, $0.5\,$, $1$ yr) and shows the corresponding SNR 
of the EMRI signal.  \label{table_RR}}
\begin{ruledtabular}
\begin{tabular}{l c c c c c c c c}  
With RR  & \multicolumn{2}{c}{$T^{}_{\rm evol}=0.1\,$yr} & \multicolumn{2}{c}{$T^{}_{\rm evol}=0.3\,$yr} & \multicolumn{2}{c}{$T^{}_{\rm evol}=0.5\,$yr} & \multicolumn{2}{c}{$T^{}_{\rm evol}=1\,$yr} \\
         & \multicolumn{2}{c}{SNR$=14.5$}                & \multicolumn{2}{c}{SNR$=43.2$}                & \multicolumn{2}{c}{SNR$=55.4$}                & \multicolumn{2}{c}{SNR$=73.5$} \\
\hline
~~~$\lambda^{i}$         & $\Delta \lambda^{i}$ & $|\log r|$             &  $\Delta \lambda^{i}$ & $|\log r|$            & $\Delta\lambda^{i}$  & $|\log r|$         &  $\Delta\lambda^{i}$ & $|\log r|$  \\
$\log{M^{}_{\bullet}}$   &  $1.4\cdot10^{-1}$   &  $1.5\cdot10^{-1}$     &  $9.2\cdot10^{-3}$    &  $1.1\cdot10^{-1}$    &  $4.5\cdot10^{-3}$   &  $1.5\cdot10^{-1}$ &  $9.3\cdot10^{-4}$   &  $2.4\cdot10^{-1}$\\
$a/M^{}_{\bullet}$       &  $1.2\cdot10^{-4}$   &  $3.4\cdot10^{-1}$     &  $1.5\cdot10^{-5}$    &  $2.0\cdot10^{-1}$    &  $4.9\cdot10^{-6}$   &  $5.3\cdot10^{-2}$ &  $1.5\cdot10^{-6}$   &  $2.3\cdot10^{-1}$\\
$e^{}_{0}$               &  $5.2\cdot10^{-6}$   &  $5.2\cdot10^{-2}$     &  $9.6\cdot10^{-7}$    &  $3.0\cdot10^{-2}$    &  $5.0\cdot10^{-7}$   &  $9.7\cdot10^{-3}$ &  $2.8\cdot10^{-7}$   &  $6.0\cdot10^{-3}$\\
$\log\,\zeta$            &  $1.1$               &  $9.3\cdot10^{-1}$     &  $1.5\cdot10^{-1}$    &  $3.1\cdot10^{-1}$    &  $4.9\cdot10^{-2}$   &  $3.1\cdot10^{-2}$ &  $2.0\cdot10^{-2}$   &  $1.5\cdot10^{-1}$\\
$\log{(D^{}_{\rm L}/\mu)}$ & $2.0\cdot10^{-1}$  &  $2.0\cdot10^{-1}$     &  $1.5\cdot10^{-1}$    &  $4.1\cdot10^{-4}$    &  $1.8\cdot10^{-2}$   &  $1.6\cdot10^{-4}$ &  $1.3\cdot10^{-2}$   &  $2.6\cdot10^{-4}$\\
\hline
\hline
With no RR       &   &   &   \multicolumn{2}{c}{$T^{}_{\rm evol}=0.3\,$yr} & \multicolumn{2}{c}{$T^{}_{\rm evol}=0.5\,$yr} & \multicolumn{2}{c}{$T^{}_{\rm evol}=1\,$yr} \\
                 &   &   &   \multicolumn{2}{c}{SNR$=38.4$}                & \multicolumn{2}{c}{SNR$=46.8$}                & \multicolumn{2}{c}{SNR$=54.6$} \\
\hline
~~~$\lambda^{i}$ &   &     & $\Delta\lambda^{i}$  &  $|\log r|$             & $\Delta\lambda^{i}$ & $|\log r|$          & $\Delta\lambda^{i}$ & $|\log r|$\\
$\log{M^{}_{\bullet}}$   &&&  $8.3\cdot10^{-3}$   &  $4.3\cdot10^{-2}$      &  $3.9\cdot10^{-3}$  &  $3.7\cdot10^{-2}$ &  $6.6\cdot10^{-4}$  &  $2.3\cdot10^{-4}$\\
$a/M^{}_{\bullet}$       &&&  $2.3\cdot10^{-5}$   &  $3.3\cdot10^{-1}$      &  $1.4\cdot10^{-5}$  &  $2.4\cdot10^{-1}$ &  $7.4\cdot10^{-6}$  &  $1.6\cdot10^{-1}$\\
$e^{}_{0}$               &&&  $1.0\cdot10^{-6}$   &  $5.0\cdot10^{-2}$      &  $6.7\cdot10^{-7}$  &  $3.6\cdot10^{-2}$ &  $1.4\cdot10^{-6}$  &  $1.6\cdot10^{-3}$\\
$\log\,\zeta$            &&&  $2.5\cdot10^{-1}$   &  $6.1\cdot10^{-1}$      &  $1.5\cdot10^{-1}$  &  $3.6\cdot10^{-1}$ &  $1.1\cdot10^{-1}$  &  $3.2\cdot10^{-1}$\\
$\log{(D^{}_{\rm L}/\mu)}$&&& $2.8\cdot10^{-2}$   &  $4.8\cdot10^{-4}$      &  $2.1\cdot10^{-2}$  &  $4.8\cdot10^{-4}$ &  $1.8\cdot10^{-2}$  &  $3.8\cdot10^{-4}$\\
\end{tabular}  
\end{ruledtabular}
\end{table*} 

We can also explore how the error estimates change with the spin parameter $a$. To that end, 
we did simulations of systems $A$ and $B$ using the following values of the spin:
$a/M^{}_{\bullet} = 0.1$ ($A^{}_1,B^{}_1$), $a/M^{}_{\bullet} = 0.25$ ($A^{}_2,B^{}_2$), and 
$a/M^{}_{\bullet} = 0.5$ ($A^{}_3,B^{}_3$). The initial semilatus rectum was set to $p^{}_0=11\,M^{}_{\bullet}$, 
which means that after evolving systems $A^{}_1$-$A^{}_3$ for a total time of $T^{}_{\rm evol}=0.5\,$yr  
and systems $B^{}_1$-$B^{}_3$ for a total time $T^{}_{\rm evol}=1.5\,$yr, the final semilatus rectum, 
$p^{}_f$, is approximately $8\,M^{}_{\bullet}$ for all systems.  
The parameter estimation errors are shown in Table~\ref{table_spins} (the upper part corresponds to
simulations of system A whereas the lower part corresponds to simulations of system B).
The first thing that we notice is that the smaller the spin parameter $a/M^{}_{\bullet}$ becomes, the better the parameter
estimate for the CS parameter $\zeta$.
In particular $\Delta \zeta\sim 2.8\cdot10^{-2}$ for system $A^{}_1$ and 
$\Delta \zeta\sim 1.4\cdot10^{-2}$ for system $B^{}_1$. 
The reason for this is quite simple.  The CS modifications affect a single MBH metric component,
${\met}^{}_{t\phi}$ [Eq.~(\ref{DCSMG_metric})], which contains the CS parameter $\xi/M_{\bullet}^4$, 
multiplied by the spin parameter $a$.  The unperturbed Kerr metric component is proportional to $a$, 
and so the relative change in this metric coefficient due to the addition of the DCSMG correction 
is proportional to $\xi$. Since we keep $\zeta= a\,\xi$ fixed as we vary $a$, the value of $\xi$ 
increases as $a$ decreases and so the CS correction to the MBH metric is larger relative to the 
leading order Kerr metric term.

The values of the SNR that we obtain for systems $A^{}_1$, $A^{}_2$, and $A_3$ are,  
$48.1$, $46.5$, and $44.7$ respectively.  In the case of systems $B^{}_1$, $B^{}_2$, and $B_3$,
the values of the SNR are $50$, $46$, and $49$. 
Notice that the SNR varies by modifying the value of the spin parameter $a$, albeit not very much.
This dependence of the SNR on the system parameters is expected in the region of the parameter space 
where the FM can be linearized (see e.g.~\cite{Nicholson:1997qh} and~\cite{Stroeer2006}). 

Overall, the parameter estimation errors for system $A$ have the magnitudes: 

\begin{eqnarray}
&& \Delta\log M^{}_{\bullet} \sim 5\cdot10^{-3}\,, \qquad \Delta a\sim 5\cdot10^{-6}\,M^{}_{\bullet}\,, \\
&& \Delta e^{}_{0}\sim 3\cdot10^{-7}\,,  ~~~~~~\qquad \Delta\log\,\zeta\sim 4\cdot10^{-2}\,, \\
&& \Delta\log( D_L/\mu)\sim 2\cdot10^{-2}\,. \label{parameter-estimation-a}
\end{eqnarray}
In the case of system $B$ they are:
\begin{eqnarray}
&& \Delta\log M^{}_{\bullet} \sim 6\cdot10^{-4}\,, \qquad \Delta a\sim 3\cdot10^{-6}\,M^{}_{\bullet}\,, \\
&& \Delta e^{}_{0}\sim 10^{-7}\,, ~~~~~~~~~~\qquad  \Delta\log\,\zeta\sim2\cdot10^{-2}\,, \\ 
&& \Delta\log( D_L/\mu)\sim 2\cdot10^{-2}\,. \label{parameter-estimation-b}
\end{eqnarray}
The order of magnitude is roughly the same for both systems, but in general the estimations for system 
$B$ are better than those for system $A$, since the MBH mass for system $B$ is larger than the one for 
system $A$ and the integration time is longer, so there are more observed waveform cycles.

\begin{table*}[ht]
\caption{Error estimates for LISA and the EMRI systems $A$ and $B$. The results shown have been 
obtained for different values of the initial eccentricity, $e^{}_0$, and MBH spin, $a$.
The evolution time for these systems is: $T^{}_{\rm evol} = 0.5\,$yr (system A) 
and $T^{}_{\rm evol} = 1.5\,$yr (system B). 
The superscript ``${}^{\dag}$'' on a given result indicates that the corresponding Fisher 
Matrix did not satisfy the MMC criterion, nevertheless we include the results for the sake of 
completeness. \label{table_spins}}
\begin{ruledtabular}
\begin{tabular}{l|ccc|ccc|ccc}  
system A                & \multicolumn{3}{c|}{$e^{}_0=0.1$}   & \multicolumn{3}{c|}{$e^{}_0=0.25$} & \multicolumn{3}{c}{$e^{}_0=0.5$}\\
\hline
$a/M^{}_{\bullet}$      &  $0.1$  &  $0.25$  &  $0.5^{\dag}$  &  $0.1$  &  $0.25$  &  $0.5$        &  $0.1$  &  $0.25$  &  $0.5$\\
\hline				
$\log{M^{}_{\bullet}}$  & $4.2\cdot10^{-3}$ & $4.1\cdot10^{-3}$ & $3.0\cdot10^{-3}$ & $3.7\cdot10^{-3}$ & $4.3\cdot10^{-3}$ & $4.4\cdot10^{-3}$ & $1.4\cdot10^{-3}$ & $5.0\cdot10^{-3}$ & $4.9\cdot10^{-3}$\\
$a/M^{}_{\bullet}$      & $5.0\cdot10^{-5}$ & $6.0\cdot10^{-6}$ & $8.0\cdot10^{-6}$ & $3.2\cdot10^{-6}$ & $5.2\cdot10^{-6}$ & $7.2\cdot10^{-6}$ & $4.0\cdot10^{-6}$ & $4.6\cdot10^{-6}$ & $6.0\cdot10^{-6}$\\
$e$                     & $2.3\cdot10^{-6}$ & $2.4\cdot10^{-6}$ & $1.4\cdot10^{-6}$ & $4.9\cdot10^{-7}$ & $8.6\cdot10^{-7}$ & $9.2\cdot10^{-7}$ & $2.0\cdot10^{-7}$ & $3.3\cdot10^{-7}$ & $3.3\cdot10^{-7}$\\
$\log{\zeta}$           & $7.5\cdot10^{-2}$ & $9.9\cdot10^{-2}$ & $9.0\cdot10^{-2}$ & $2.8\cdot10^{-2}$ & $4.9\cdot10^{-2}$ & $6.6\cdot10^{-2}$ & $5.1\cdot10^{-2}$ & $3.5\cdot10^{-2}$ & $4.3\cdot10^{-2}$\\
$\log{(D^{}_L/\mu)}$    & $1.9\cdot10^{-2}$ & $2.0\cdot10^{-2}$ & $2.1\cdot10^{-2}$ & $1.7\cdot10^{-2}$ & $2.0\cdot10^{-2}$ & $2.1\cdot10^{-2}$ & $1.9\cdot10^{-2}$ & $2.3\cdot10^{-2}$ & $2.4\cdot10^{-2}$\\ 
\hline\hline
system B                & \multicolumn{3}{c|}{$e^{}_0=0.1$}   & \multicolumn{3}{c|}{$e^{}_0=0.25$} & \multicolumn{3}{c}{$e^{}_0=0.5$}\\
\hline
$a/M^{}_{\bullet}$      &  $0.1$  &  $0.25$  &  $0.5^{\dag}$  &  $0.1$  &  $0.25$  &  $0.5$        &  $0.1$  &  $0.25$  &  $0.5$\\
\hline				
$\log{M^{}_{\bullet}}$  & $1.1\cdot10^{-3}$ & $1.1\cdot10^{-3}$ & $5.3\cdot10^{-4}$ & $8.7\cdot10^{-4}$ & $9.0\cdot10^{-4}$ & $9.9\cdot10^{-4}$ & $6.1\cdot10^{-4}$ & $6.4\cdot10^{-4}$ & $6.8\cdot10^{-4}$\\
$a/M^{}_{\bullet}$      & $4.8\cdot10^{-6}$ & $7.7\cdot10^{-6}$ & $5.1\cdot10^{-6}$ & $3.2\cdot10^{-6}$ & $4.3\cdot10^{-6}$ & $5.2\cdot10^{-6}$ & $1.7\cdot10^{-6}$ & $2.1\cdot10^{-6}$ & $2.7\cdot10^{-6}$\\
$e$                     & $1.5\cdot10^{-6}$ & $1.9\cdot10^{-6}$ & $6.0\cdot10^{-7}$ & $4.1\cdot10^{-7}$ & $4.3\cdot10^{-7}$ & $4.5\cdot10^{-7}$ & $9.6\cdot10^{-8}$ & $1.0\cdot10^{-7}$ & $1.1\cdot10^{-7}$\\
$\log{\zeta}$           & $6.4\cdot10^{-2}$ & $8.5\cdot10^{-2}$ & $7.0\cdot10^{-2}$ & $3.7\cdot10^{-2}$ & $4.3\cdot10^{-2}$ & $5.3\cdot10^{-2}$ & $1.3\cdot10^{-2}$ & $1.6\cdot10^{-2}$ & $2.1\cdot10^{-2}$\\
$\log{(D^{}_L/\mu)}$    & $2.6\cdot10^{-2}$ & $2.7\cdot10^{-2}$ & $2.0\cdot10^{-2}$ & $2.3\cdot10^{-2}$ & $2.4\cdot10^{-2}$ & $2.6\cdot10^{-2}$ & $2.0\cdot10^{-2}$ & $2.1\cdot10^{-2}$ & $2.2\cdot10^{-2}$\\ 
\end{tabular} 
\end{ruledtabular}
\end{table*} 

The parameter error estimates presented are for a fixed value of the parameter $\zeta=\xi\cdot a$. 
Since the spin parameter $a/M^{}_{\bullet}$ is fixed and is the same for both systems $A$ and $B$ 
in Table~\ref{EMRI_systems}, this means that in the previous results the CS parameter $\xi/M_{\bullet}^4$ 
was fixed. Now, we present results for the EMRI system $A$ for different values of the CS parameter 
$\xi/M_{\bullet}^4$.  We have considered the following particular values: $\xi=0.05M_{\bullet}^4$, 
$\xi=0.1M_{\bullet}^4$ and $\xi=0.2M_{\bullet}^4$.  The results obtained for the estimation of 
the parameter errors of $\mb{\lambda} = \{M^{}_{\bullet}, a/M^{}_{\bullet}, e^{}_0, p^{}_0, \zeta, D^{}_{\rm L}/\mu\}\,$ 
are shown in Table~\ref{table_diff_xis}.
Due to the fact that the dependence on $\xi/M_{\bullet}^4$ and on $a/M^{}_{\bullet}$ are different in the MBH metric components 
and in the evolution equations (see Sec.~\ref{formulation}), one would expect a different dependence 
of the error estimates when varying both parameters independently.  By comparing the results of
Tables~\ref{table_spins} and~\ref{table_diff_xis}, we can see that modifying the value of the CS 
parameter $\xi/M_{\bullet}^4$ only affects significantly the error estimate of the CS parameter itself, whereas 
modifying $a/M^{}_{\bullet}$ has a significant effect on the error estimates of all the parameters employed in our study, 
and in particular on $\zeta$.

\begin{table}
\caption{Error estimates for LISA and the EMRI system $A$ in Table~\ref{EMRI_systems} obtained
by changing the value of the CS parameter $\xi$. 
As we can see, by increasing the value of the CS parameter, $\xi$, its error estimate, 
$\Delta \log{\zeta}$, improves, whereas the rest of the error 
estimates remain roughly constant. \label{table_diff_xis} } 
\begin{ruledtabular}
\begin{tabular}{lccc}  
                        & $\xi/M_{\bullet}^4=0.05$ & $\xi/M_{\bullet}^4=0.1$ & $\xi/M_{\bullet}^4=0.2$  \\
\hline
$\log{M^{}_{\bullet}}$  & $4.4\cdot10^{-3}$        & $4.2\cdot10^{-3}$ & $4.5\cdot10^{-3}$\\
$a/M^{}_{\bullet}$      & $4.9\cdot10^{-6}$ & $4.7\cdot10^{-6}$ & $4.9\cdot10^{-6}$\\
$e^{}_{0}$              & $4.9\cdot10^{-7}$ & $4.9\cdot10^{-7}$ & $5.0\cdot10^{-7}$\\
$\log{\zeta}$           & $1.9\cdot10^{-1}$ & $9.5\cdot10^{-2}$ & $4.9\cdot10^{-2}$ \\
$\log{(D^{}_L/\mu)}$    & $1.8\cdot10^{-2}$ & $1.8\cdot10^{-2}$ & $1.8\cdot10^{-2}$\\
\end{tabular} 
\end{ruledtabular}
\end{table} 

\begin{table}
\caption{Parameter estimation results for eLISA/NGO and the EMRI System $A$. 
evolved for $T^{}_{\rm evol} = 2\,$yr.  The corresponding SNR is $\simeq 15\,$.
The value of the initial eccentricity is $e^{}_0= 0.25\,$. \label{table_eLISA}}  
\begin{ruledtabular}
\begin{tabular}{l c c c} 
                       & $a/M^{}_{\bullet} = 0.1$ & $a/M^{}_{\bullet} = 0.25$ & $a/M^{}_{\bullet} = 0.5$  \\
\hline
$\log{M^{}_{\bullet}}$ & $9.0\cdot10^{-4}$        & $9.8\cdot10^{-4}$         & $1.3\cdot10^{-3}$\\
$a/M^{}_{\bullet}$     & $3.2\cdot10^{-6}$        & $2.8\cdot10^{-6}$         & $3.9\cdot10^{-6}$\\
$e^{}_{0}$             & $5.1\cdot10^{-7}$        & $5.2\cdot10^{-7}$         & $5.7\cdot10^{-7}$\\
$\log{\zeta}$          & $6.0\cdot10^{-2}$        & $7.3\cdot10^{-2}$         & $9.6\cdot10^{-2}$ \\
$\log{(D^{}_L/\mu)}$   & $6.4\cdot10^{-2}$        & $7.0\cdot10^{-2}$         & $7.5\cdot10^{-2}$\\
\end{tabular} 
\end{ruledtabular}
\end{table} 

Up to now, all the parameter estimation results presented refer to the LISA detector.
We now present some results for eLISA/NGO~\cite{AmaroSeoane:2012km}.  
In order to more easily compare with the results obtained for LISA, we normalize to a fixed 
SNR, since the SNR for eLISA/NGO is around two times smaller. 
We considered system A with three different values of the spin parameter $a/M^{}_{\bullet}$ namely
$a/M^{}_{\bullet} = 0.1\,$, $0.25\,$, and $0.5\,$. 
The results obtained are quoted in Table~\ref{table_eLISA}. Comparing them with the ones quoted in 
Table~\ref{table_spins} for LISA, we can see that the parameter estimation accuracy 
does not change appreciably when the noise curve of LISA is changed for the one of 
eLISA/NGO and so, all previous results can be considered to apply to eLISA/NGO as well, 
with the corresponding SNR corrections.

\begin{table*}
\centering   
\caption{Parameter estimation errors for EMRI systems in GR (i.e. on the $\mb{\lambda}$-parameter surface
determined by $\xi = \zeta = 0$).  The parameters common to all these systems are: 
$M^{}_{\bullet}= 5\cdot 10^{5} M^{}_{\odot}\,$, 
$\mu = 2\cdot10^{-5}\,$,
$\theta^{}_{\rm S} = \phi^{}_{\rm S} = 1.57\,$rad,
$\theta^{}_{\rm K} = 0.392\,$rad,
$\phi^{}_{\rm K} = 0.78\,$rad,
$D^{}_{\rm L}/\mu = 5\cdot10^{4}\,$Gpc, and
$\psi^{}_{0} = \chi^{}_{0} = \phi^{}_{0}= 0.78\,$rad. 
The last two columns contain the error estimate for the CS parameter $\xi$ and the MMC criterium figure
of merit associated with the CS parameter $\zeta\,.$
\label{boundxitable}} 
\begin{ruledtabular}
\begin{tabular}{ccccc|cc}
$a/M^{}_{\bullet}$  & $e^{}_{0}$ & $p^{}_{0}$ & $\theta^{}_{\INC,0}$ & $T^{}_{\rm evol}$ (yrs) & $\Delta\xi/M^{4}_{\bullet}$ & $|\log r|^{}$ \\[1mm]
\hline
$0.5$               & $0.7$      & $10.0$     & $0.15$               & $0.5$                   & $5.76\cdot 10^{-8}$         & $0.91$ \\[1mm]         
$0.45$              & $0.7$      & $10.0$     & $0.15$               & $0.5$                   & $1.86\cdot 10^{-7}$         & $0.84$ \\[1mm]
$0.5$               & $0.7$      & $10.0$     & $0.15$               & $0.5$                   & $6.23\cdot 10^{-8}$         & $0.89$ \\[1mm]
$0.5$               & $0.85$     & $11.0$     & $0.15$               & $0.5$                   & $6.20\cdot 10^{-8}$         & $0.7$ \\[1mm]
$0.5$               & $0.85$     & $11.0$     & $0.15$               & $1.0$                   & $6.10\cdot 10^{-8}$         & $0.57$
\end{tabular} 
\end{ruledtabular}
\end{table*}

\section{Placing a Bound on the CS Parameter}\label{Bounds}

One application of the framework we have developed to perform parameter error studies in
DCSMG is to try to put bounds on the CS parameter $\xi$, which is the combination of
CS coupling constants and the gravitational constant that controls deviations from GR
in the dynamics of EMRIs.  This question has already been investigated
in the literature, but using astrophysical systems different from EMRIs.

In the case of non-dynamical CS gravity, although the scalar field $\vartheta$ has no 
evolution equation, it can be prescribed a certain time evolution that has an associated 
time-derivative $\dot{\vartheta}$ and timescale, $\tau^{}_{\rm CS}\, = 1/\dot{\vartheta}$. 
Bounds are normally written in terms of constraints on $\ell^2/\tau^{}_{\rm CS}$, where 
$\ell^2$ is the characteristic length scale and equals the coupling constant $\alpha\,$ introduced earlier. 
Strong bounds on this combination were first obtained by Yunes and Spergel~\cite{Yunes:2008ua}
but refinements introduced by Ali-Haimoud~\cite{AliHaimoud:2011bk} set the bound to
$0.2\,$km, which is three orders of magnitude better than the Solar System bound~\cite{Smith:2007jm}, 
obtained from data from the LAGEOS satellites orbiting the Earth~\cite{Ciufolini:2004rq}.

For dynamical Chern-Simons gravity, which we consider here, the bound is normally expressed 
as a bound on $\xi^{1/4}\,$. The first bound was quoted by Yunes and Pretorius~\cite{Yunes2009}
and was $\xi^{1/4} < 10^4\,$km.  However, in a recent paper by Ali-Haimoud and Chen~\cite{AliHaimoud:2011fw}
they took into account the fact that the CS solution for the spacetime outside a rotating star 
is not the same as that outside a rotating black hole, and also that the CS correction can only lead 
to a decrease in frame-dragging effects and thus cannot be constrained by an upper bound 
on the precession, but only by a positive lower bound that lies below the GR value.  The CS-induced 
precession that gives the bound quoted in Yunes and Pretorius is two orders of magnitude 
larger than the GR precession, which means that bound can probably not be trusted. 
Ali-Haimoud and Chen~\cite{AliHaimoud:2011fw} argue that the best constraints at present are 
therefore those that come from Solar System measurements, based on data
from the Gravity Probe B satellite~\cite{Everitt:2011hp} and also from the LAGEOS 
satellites~\cite{Ciufolini:2004rq}, which are much weaker.  The bound that they get is 
then $\xi^{1/4} < 10^8\,$km.   In this paper we compare 
our results with this weaker but more robust bound.  

The basis for the computation of our bound is the following.  We assume that
GR is the correct theory to describe EMRI dynamics and hence assume that measurements made
by LISA are compatible with $\xi=0$.  Then, by estimating the error on the measurements of 
$\xi$, $\Delta\xi$ [obtained using Eq.~(\ref{estimated-error})], we can set a bound of the 
following type: $\xi < \Delta\xi\,$.  Different EMRI systems will provide different constraints.  
But since $\xi$ is a universal quantity, in particular the same for all EMRIs,
we just need to look for the EMRI system that provides the best constraint. We have performed 
several computations with EMRI systems whose common parameters are: 
$M^{}_{\bullet}= 5\cdot 10^{5} M^{}_{\odot}\,$, $\mu = 2\cdot10^{-5}\,$, $\theta^{}_{\rm S} = 
\phi^{}_{\rm S} = 1.57\,$rad, $\theta^{}_{\rm K} = 0.392\,$rad,
$\phi^{}_{\rm K} = 0.78\,$rad, $D^{}_{\rm L}/\mu = 5\cdot10^{4}\,$Gpc, and $\psi^{}_{0} = 
\chi^{}_{0} = \phi^{}_{0}= 0.78\,$rad. We show some relevant results in Table~\ref{boundxitable} 
for EMRIs with  spin parameter $a/M^{}_{\bullet} \approx 0.5$ (the rest of parameters can be 
found in the caption of the table). Since we are differentiating about zero, the numerical 
evaluation of  the $\xi$ derivatives must be performed using a one-sided derivative. 
In particular, we have double-checked some of these results using both the 3-point and 
4-point rules given by Eqs.~(\ref{threepointrule}) and~(\ref{fourpointrule}) respectively. 
We note that, even though we are using system $A$ ($M_{\bullet} = 5\cdot10^5M_{\odot}$) for 
this study, the values obtained for the MMC with $T_{\rm evol} = 0.5$ were slightly above the 
reference threshold of $0.5$ that we used elsewhere in our study. This fact could be connected 
to using the one-sided derivative in our calculations. 

From the error estimates for the $\xi$ parameter shown in Table~\ref{boundxitable} , we find
\begin{equation}
\Delta\xi/M^{4}_{\bullet} < 10^{-7}\,.
\end{equation}
which, in suitable units, becomes
\begin{equation}
\xi^{1/4} < 1.4\cdot 10^{4}{\rm km}\,.
\end{equation}
This result, a prediction for LISA measurements, is almost four orders of magnitude better 
than the bound $\xi^{1/4}\lesssim 10^8\,$km given in~\cite{AliHaimoud:2011fw}.  The 
corresponding estimate for eLISA/NGO can be found be rescaling the $\xi$ constraint by the 
SNR, but since the bound scales only as the one-fourth power, the bound for eLISA/NGO 
is essentially the same.

\section{Conclusions and Discussion}\label{conclusions}
In this paper we have examined how well a space-based GW detector like LISA 
or eLISA/NGO can discriminate between an EMRI system in GR and one occurring in a modified gravity 
theory like DCSMG. To do this, we have extended previous work in~\cite{Sopuerta:2009iy} 
by introducing two key components.  The first is the inclusion of RR effects driving the inspiral. 
We have constructed a waveform template model using an adiabatic-radiative approximation
following the NK waveform model~\cite{Babak:2006uv}.
In this approximation, the inspiral trajectory is modeled as a sequence of geodesics whose constants of motion
are updated using formulae for the fluxes of energy, $z$-component of the angular momentum,
and Carter constant, that were derived for general relativistic inspirals in~\cite{Gair2006} 
using a combination of PN approximations
and fits to results from the Teukolsky formalism. 

The second key improvement made in this paper is the use of the Fisher matrix formalism to
estimate errors in parameter measurements. We have explored a five-dimensional subspace
of the fifteen-dimensional parameter space of EMRIs in DCSMG (see Table~\ref{parameters}). 
The parameters that span this subspace are $\{M^{}_{\bullet}, a/M^{}_{\bullet} , e^{}_0, \zeta, D^{}_L/\mu\}$.
We have focused our studies on two types of systems (see Table~\ref{EMRI_systems}),
with masses $10M^{}_{\odot}+5\cdot10^{5}M^{}_{\odot}$ and $10M^{}_{\odot}+10^{6}M^{}_{\odot}\,$. 
The parameter error estimates are summarized in Eqs.~(\ref{parameter-estimation-a}) 
and~(\ref{parameter-estimation-b}).  For both systems, and assuming a LISA detector, 
we estimated the measurement error on the logarithm of the CS parameter $\zeta$ as $\Delta\log\zeta\sim10^{-2}$. 
Therefore, a space-based GW detector like LISA should be able to discriminate between 
GR and DCSMG. In the case that DCSMG is the correct theory describing the strong gravitational
regimes involved in EMRI dynamics, such a detector should be able to provide a good
estimation of the CS parameter that controls the deviations from GR.  We have also
explored how these parameter error estimates change with the spin parameter $a/M^{}_{\bullet} $ (see
Table~\ref{table_spins}).  We have found that by decreasing $a/M^{}_{\bullet} $ the parameter measurement 
precision of the CS parameter improves, while modifying the value of $\xi$ (see Table~\ref{table_diff_xis})
does not affect significantly the precision of parameter estimates for the range of other system parameters 
included in our analysis.  

For the case of eLISA/NGO we have presented some parameter error estimation results for system 
$A$ in Table~\ref{EMRI_systems}.  In order to compare with LISA results we have
normalized these results to a fixed SNR (the eLISA/NGO SNR is approximately a factor of two smaller
than the LISA one).  Results for three values of the spin parameter 
($a/M^{}_{\bullet} = 0.1\,$, $0.25\,$, and $0.5\,$) are given in Table~\ref{table_eLISA}.
The conclusion is that the parameter estimation accuracy at fixed SNR does not change significantly
relative to the LISA results. The LISA results can therefore be applied to eLISA/NGO by
applying the appropriate SNR correction.

Finally, we have used our parameter estimation framework to put bounds on the CS
parameter $\xi$.  By assuming that GR is the correct theory of gravity we have found 
that LISA could place a bound $\xi^{1/4} <  1.4\cdot10^4\,$km,
which is almost four orders of magnitude better than the bound obtained 
in~\cite{AliHaimoud:2011fw} using Solar System data.

The results presented in this paper can be extended in a number of ways by adding
more elements to the waveform model that we employ. For example it can be done by:
(i) Using a higher order approximation for the MBH geometry in DCSMG;  (ii) including CS corrections to
the RR formulae, in particular to introduce the effects of the CS scalar field 
in the RR mechanism;  (iii) adding more multipole moments to the gravitational 
wave expansion formulae;  etc.  In addition, we have focussed our study on
a few EMRI systems, so it would be useful to carry out a more exhaustive study of the parameter
space, although this would be a costly task in terms of computational resources.  
Such extensions to the present work would allow
us to consider systems that might be of greater interest from the point of view of improving
the parameter estimation results.  The approximations underlying our model prevent
us from considering systems with spins higher than $a/M^{}_{\bullet}=0.5$ and
strong CS couplings.   However, a better search of the parameter space
would allow us to identify systems that provide the best parameter estimates and the strongest 
bounds on the CS parameter $\xi$.

There are other extensions of this work that are also interesting.  In particular, 
it would be useful to assess the systematic errors that would arise if GR waveform templates were used to 
detect EMRIs that are actually described by DCSMG. This could be done using the 
formalism developed by Cutler and Vallisneri~\cite{Cutler:2007mi} to estimate 
systematic errors that arise from model uncertainties.  Finally, we could apply
some of the tools and techniques used in the present work to study other modifications of
gravity, different from the CS correction and in this way to exploit the potential 
of the connection between gravitational wave astronomy and high-energy physics~\cite{Cardoso:2012qm}.

\section*{Acknowledgments}
We would like to thank Leor Barack, Edward K. Porter, Michele Vallisneri, Kent Yagi, and Nicol\'as Yunes for helpful 
discussions.  PCM work has been supported by a predoctoral FPU fellowship of the Spanish 
Ministry of Science and Innovation (MICINN) and by the Beatriu de Pin\'os programme of the 
Catalan Agency for Research Funding (AGAUR).  JG's work is supported by the Royal Society.
CFS acknowledges support from the Ram\'on y Cajal Programme of the Ministry of Education 
and Science of Spain, by a Marie Curie International Reintegration Grant (MIRG-CT-2007-205005/PHY) 
within the 7th European Community Framework Programme, from contract 2009-SGR-935 of AGAUR,
and from contracts FIS2008-06078-C03-03, AYA-2010-15709, and FIS2011-30145-C03-03 of MICCIN. 
We acknowledge the computational resources provided by the Barcelona Supercomputing Centre
(AECT-2011-3-0007) and CESGA (contracts CESGA-ICTS-200 and CESGA-ICTS-221).

\appendix

\section{LISA and eLISA Power Spectral Densities}\label{lisaspds}

In this paper we assume that the GW detector is either LISA~\cite{Danzmann:2003tv,Prince:2003aa} 
or eLISA/NGO~\cite{AmaroSeoane:2012km}.  LISA is a space-based GW detector concept that consists
of a  quasi-equilateral triangular constellation of three identical spacecrafts with an 
inter-spacecraft distance of $L=5\cdot 10^{9}\,$m.  Each spacecraft follows a heliocentric
orbit that trails behind the Earth at a distance of $5\cdot10^{10}\,$m (equivalent to $20$ degrees)
in such a way that the LISA constellation faces the Sun, slanting at $60$ degrees to the ecliptic plane. 
These particular heliocentric orbits were chosen such that the triangular formation is maintained 
throughout the year, with the triangle appearing to rotate about the center of the formation once per year. 
Each spacecraft contains two free-falling test masses whose distance is monitored by $6$ laser links.
In contrast, the eLISA/NGO constellation has an inter-spacecraft distance of $L=10^9$m. Moreover, only 
one of the spacecrafts will contain two free falling masses and service two arms 
of the constellation, while the other two will have only one proof mass and service one arm. 
This effectively reduces the detector response from having two independent Michelson channels to just one. 

An essential ingredient required in the detector response is a model for the noise affecting 
the observations. This may be described in terms of the one-sided noise power spectral 
density, $S^{}_{n}(f)$. For LISA, this has three contributions: instrumental noise, 
$S_{n}^{\text{inst}}(f)$, confusion noise from short-period galactic binaries, 
$S_{n}^{\text{gal}}(f)$, and confusion noise from extragalactic binaries, 
$S_{n}^{\text{exgal}}(f)$~\cite{Barack:2003fp}:
\begin{eqnarray}
S^{}_n = \text{min}\left\{ S_{n}^{\text{inst}}+  S_{n}^{\text{exgal}}\ , \  
S_n^{\text{inst}}+ S_{n}^{\text{gal}}+ S_{n}^{\text{exgal}}\right\}\,,\label{ps_LISA}
\end{eqnarray}
where the different noise contributions are given by:
\begin{eqnarray}
S_{n}^{\text{inst}}(f) &=& \exp{\left(\kappa T^{-1}_{\text{mission}}
\frac{{\rm d}N}{{\rm d}f}\right)} (9.18\times10^{-52}f^{-4} \nonumber \\ 
& + & 1.59\times10^{-41} + 9.18\times10^{-38}f^{2})\; \text{Hz}^{-1},~~~~~\\[2mm]
S_{n}^{\text{gal}}(f) &=& 2.1\times10^{-45}
\left(\frac{f}{1\text{Hz}}\right)^{-7/3}\; \text{Hz}^{-1}\,,\\[2mm]
S_{n}^{\text{exgal}}(f) &=& 4.2\times10^{-47}
\left(\frac{f}{1\text{Hz}}\right)^{-7/3}\; \text{Hz}^{-1}\,,
\end{eqnarray}
with ${\rm d}N/{\rm d}f $ the number density of galactic white dwarf binaries per unit 
frequency, $T_{\rm mission}$  the lifetime of the LISA mission, and $\kappa$ the average 
number of frequency bins that are lost when each galactic binary is fitted out. 
The particular values that we use correspond to (see, e.g.~\cite{Barack:2003fp}) 
$\kappa \approx 4.5$ and:
\begin{eqnarray}
\frac{{\rm d}N}{{\rm d}f} = 2\times 10^{-3}\left( \frac{1\text{Hz}}{f}\right)^{11/3}\,.
\end{eqnarray}

The  noise curve for eLISA/NGO is given by~\cite{AmaroSeoane:2012km}:
\begin{eqnarray}
S^{}_{n}(f) = \frac{4S^{}_{\text{acc}} + S^{}_{\text{sn}} + S^{}_{\text{omn}}}{L^2}
\left[1+\left( \frac{fL}{0.205c}\right)^2\right]\,, 
\end{eqnarray}
where $S^{}_{\text{acc}}$, $S^{}_{\text{sn}}$ and $S^{}_{\text{omn}}$ are, respectively, 
the power spectral density of the residual acceleration of the test masses, of the shot 
noise and of other measurement noises.  These are given by:
\begin{eqnarray}
S^{}_{\text{acc}}(f) &=&1.37\cdot10^{-32}\left(1+\frac{10^{-4}\text{Hz}}{f}\right)\frac{\text{Hz}}
{f^4}\text{m}^2\text{Hz}^{-1}\,, \\
S^{}_{\text{sn}}(f) &=& 5.25\cdot10^{-23}\text{m}^2\text{Hz}^{-1}\,,\\
S^{}_{\text{omn}} &=& 6.28\cdot10^{-23}\text{m}^2\text{Hz}^{-1}\,.
\end{eqnarray}
%

\section{Evolution of the Constants of Motion}\label{evolconstantsofmotion}
In~\cite{Gair2006,Babak:2006uv}, to compute EMRIs in general relativity, the fluxes on the 
right-hand sides of Eqs.~(\ref{edot_eq})-(\ref{qdot_eq}) were specified by approximate, weak-field, formulae, 
augmented with corrections to ensure the behavior was not pathological for near-circular 
or near-polar orbits and augmented by fits to numerical solutions of the Teukolsky equation.
These formulae look as follows (note that in this paper we are using a dimensionless
semilatus rectum instead of the semilatus rectum of~\cite{Gair2006}, which has units of $M^{}_{\bullet}$):
\begin{widetext}
\begin{eqnarray}
\frac{{\rm d}E}{dt} &=& (1-e^2)^{3/2} \left[ (1-e^2)^{-3/2}(\dot{E})^{}_{\rm 2PN}(p,\iota,e,a) 
- (\dot{E})^{}_{\rm 2PN} (p,\iota,0,a) - \frac{N^{}_4(p,\iota)}{N^{}_1(p,\iota)}\,
(\dot{L}^{}_z)^{}_{\rm 2PN}(p,\iota,0,a) \right. \nonumber \\
&-&\left. \frac{N^{}_5(p,\iota)}{N^{}_1(p,\iota)} \,(\dot{Q})^{}_{\rm 2PN}(p,\iota,0,a)\right]\,, \label{Edotcorr} \\[4mm]
\frac{{\rm d}{L}^{}_z}{dt} & = & (1-e^2)^{3/2} \left[(1-e^2)^{-3/2}
(\dot{L}^{}_z)^{}_{\rm 2PN}(p,\iota,e,a) - (\dot{L}^{}_z)^{}_{\rm 2PN}(p,\iota,0,a) 
+ (\dot{L}^{}_z)^{}_{\rm fit}\right] \,, \label{Ldotfinal} \\[4mm]
\frac{{\rm d}Q}{dt} &=& (1-e^2)^{3/2} \sqrt{Q(p,\iota,e,a)} \left[(1-e^2)^{-3/2}
\left(\frac{\dot{Q}}{\sqrt{Q}}\right)^{}_{\rm 2PN}(p,\iota,e,a) 
- \left(\frac{\dot{Q}}{\sqrt{Q}}\right)^{}_{\rm 2PN}(p,\iota,0,a) \right. \nonumber \\ 
&+&\left. 2\,\tan\iota\left\{(\dot{L}^{}_z)^{}_{\rm fit} 
+\frac{\sqrt{Q(p,\iota,0,a)}}{\sin^2\iota}(\dot{\iota})^{}_{\rm fit}\right\}\right] \,.
\label{Qdotfinal}
\end{eqnarray}
\end{widetext}
where the coefficients $N^{}_i$'s are
\begin{eqnarray}
N^{}_1(p,\iota) & = &  \left. pM^{4}_{\bullet}\left[ p\,E\left(p^{2}+q^{2}\right)
-2q\left(\frac{L^{}_{z}}{M^{}_{\bullet}}-qE\right)\right] \right|^{}_{\rm circ}  \,, \\
N^{}_4(p,\iota) & = & \left. pM^{3}_{\bullet}\left[(2-p)\frac{L^{}_{z}}{M^{}_{\bullet}}
-2qE\right] \right|^{}_{\rm circ}  \,,\\ 
N^{}_5(p,\iota) & = & \left. \frac{M^{2}_{\bullet}}{2}\left[p(2-p)-q^{2} \right]
\right|^{}_{\rm circ}\,,
\label{QNdefs}
\end{eqnarray}
where the subscript `circ' indicated that these coefficients are evaluated for a circular orbit 
defined by the arguments $p$ and $\iota$.  In these expressions, $q$ denotes the dimensionless spin parameter of the MBH
\begin{equation}
q = \frac{a}{M^{}_{\bullet}}\,,\qquad
0\leq q \leq 1 \,.
\end{equation}
In Eqs.~(\ref{Edotcorr})-(\ref{Qdotfinal}), the fluxes $(\dot{E})^{}_{\rm 2PN}$, 
$(\dot{L}^{}_z)^{}_{\rm 2PN}$, and $(\dot{Q})^{}_{\rm 2PN}$ are the 2PN approximations
to the averaged evolution of the energy, angular momentum in the spin direction, and
Carter constant.  They are modifications of the original expressions given in~\cite{Glampedakis:2002cb}
but corrected to avoid unphysical features that they exhibit for nearly circular ($e\approx  0$) 
and for nearly polar ($\iota\approx\pi/2$) inspirals.
The corrected 2PN fluxes have the following form
\begin{widetext}
\begin{eqnarray}
(\dot{E})^{}_{\rm 2PN} &=& -\frac{32}{5} \frac{m^2_{\star}}{M^2_{\bullet}} 
\frac{(1-e^2)^{3/2}}{p^{5}} \left[ 
g^{}_1(e) - \frac{q}{p^{3/2}}\,g^{}_2(e)\cos\iota - \frac{1}{p}\,g^{}_3(e) 
+ \frac{\pi}{p^{3/2}}\,g^{}_4(e) - \frac{1}{p^2}\,g^{}_5(e)  
+ \frac{q^2}{p^2}\,g^{}_6(e) \right.
\nonumber \\
&-& \left. \left(\frac{527}{96} + \frac{6533}{192}e^{2} \right) 
\frac{q^2}{p^2}\,\sin^2\iota  \right ] \,,  \label{new_Edot_2}  \\[2mm]
(\dot{L}^{}_z)^{}_{\rm 2PN} &=& -\frac{32}{5} \frac{m^2_{\star}}{M^{}_{\bullet}} 
\frac{(1-e^2)^{3/2}}{p^{7/2}}\left[ g^{}_9(e)\cos\iota  
+ \frac{q}{p^{3/2}}\left\{g_{10}^{a}(e) - g_{10}^{b}(e) \cos^2\iota\right\} 
-\frac{1}{p}\,g^{}_{11}(e) \cos\iota 
\right. \nonumber \\
&+& \left. \frac{\pi}{p^{3/2}}\,g^{}_{12}(e)\cos\iota 
-\frac{1}{p^2}\,g^{}_{13}(e)\cos\iota  
+\frac{q^2}{p^2}\,\cos\iota \left\{g^{}_{14}(e) 
- \left(\frac{45}{8} +\frac{37}{2}e^{2}\right)\,\sin^2\iota\right\}  \right ] \,, 
\label{new_Ldot_2} \\[2mm]
(\dot{Q})^{}_{\rm 2PN} &=&  -\frac{64}{5} \frac{m^2_{\star}}{M^{}_{\bullet}} 
\frac{(1-e^2)^{3/2}}{p^{7/2}}\,\sqrt{Q}\,\sin\iota\,\left[ g^{}_9(e) 
-\frac{q}{p^{3/2}}\,g_{10}^{b}(e)\cos\iota
-\frac{1}{p}\,g^{}_{11}(e) + \frac{\pi}{p^{3/2}}\,g^{}_{12}(e) 
\right. \nonumber \\ 
&-& \left. \frac{1}{p^2}\,g^{}_{13}(e) + \frac{q^2}{p^2}\,\left(g^{}_{14}(e) 
- \frac{45}{8}\,\sin^2\iota\right)  \right] \,,\label{Qdot_new}
\end{eqnarray}
\end{widetext}
where the various $e$-dependent coefficients are
\begin{widetext}
\begin{eqnarray}
g^{}_1(e) &=& 1 + \frac{73}{24} e^2  + \frac{37}{96}e^4, \qquad
g^{}_2(e)= \frac{73}{12} + \frac{823}{24} e^2 + \frac{949}{32}e^4
+ \frac{491}{192}e^6 \,, \qquad
g^{}_3(e) = \frac{1247}{336} + \frac{9181}{672} e^2 \,,  
\nonumber \\[2mm] 
g^{}_4(e) &=& 4 + \frac{1375}{48} e^2  \,, \qquad
g^{}_5(e) = \frac{44711}{9072} + \frac{172157}{2592} e^2 \,,\qquad
g^{}_6(e) = \frac{33}{16} + \frac{359}{32} e^2  \,,
\nonumber \\[2mm]
g^{}_7(e) &=& \frac{8191}{672} + \frac{44531}{336} e^2 \,,\qquad
g^{}_8(e) = \frac{3749}{336} - \frac{5143}{168} e^2  \,, \qquad
g^{}_9(e) = 1 + \frac{7}{8} e^2 \,,
\nonumber \\[2mm]
g_{10}^{a}(e) & = & \frac{61}{24} + \frac{63}{8}e^2   + \frac{95}{64}e^4 \,,
\qquad 
g_{10}^{b}(e) = \frac{61}{8} + \frac{91}{4}e^2  + \frac{461}{64}e^4 \,,\qquad
g^{}_{11}(e) = \frac{1247}{336} + \frac{425}{336} e^2  \,,
\nonumber \\[2mm]
g^{}_{12}(e) & = & 4 + \frac{97}{8} e^2 \,, \qquad
g^{}_{13}(e) = \frac{44711}{9072} + \frac{302893}{6048} e^2 ,\qquad
g^{}_{14}(e) = \frac{33}{16} + \frac{95}{16} e^2 \,.
\end{eqnarray}
\end{widetext}
The equation for the evolution for the Carter constant has an additional improvement
with respect to the one in~\cite{Glampedakis:2002cb}, where a simple but accurate
prescription for the Carter constant was given by assuming that the inclination angle
evolution due to GW emission is negligible (see~\cite{Hughes:1999bq,Hughes:2001jr} for
supporting evidence of this).
That is, $\dot{\iota}\approx 0$ leads to $\dot{Q} \approx 2(\dot{L}^{}_{z}/L^{}_{z}) Q$
via Eq.~(\ref{iota-angle}).  The improvement introduced in~\cite{Gair2006} consists of adding the 
next-order spin-dependent PN correction.

The final ingredient comes by adding fitting functions to the results of Teukolsky-based 
computations for circular-inclined orbits~\cite{Hughes:1999bq}.
The expressions for the Teukolsky fitted fluxes (to data provided by Scott Hughes) are
\begin{widetext}
\begin{eqnarray}
(\dot{L}^{}_z)^{}_{\rm fit} &=& -\frac{32}{5} \frac{m^2_{\star}}{M^{}_{\bullet}}\,{p}^{-7/2}\,
\left [ \cos\iota  + \frac{q}{p^{3/2}}\left(\frac{61}{24}-\frac{61}{8}\,\cos^2\iota\right) 
-\frac{1247}{336\,p}\,\cos\iota  + \frac{4\pi}{p^{3/2}}\,\cos\iota 
- \frac{44711}{9072\,p^{2}}\,\cos\iota \right.\nonumber \\ 
&+& \left.  \frac{q^2}{p^2}\,\cos\iota\left(\frac{33}{16} - \frac{45}{8}\,\sin^2\iota\right) 
 + \frac{1}{p^{5/2}}\left\{ q\, \left(d_1^a + \frac{d^b_1}{p^{1/2}} + \frac{d^c_1}{p}\,\right) 
 + q^3\,\left(d_2^a + \frac{d^b_2}{p^{1/2}} + \frac{d^c_2}{p}\,\right) \right.\right.\nonumber \\ 
&+& \left. \cos\iota\left(c_1^a + \frac{c^b_1}{p^{1/2}} + \frac{c^c_1}{p}\,\right) 
+ q^2\,\cos\iota\,\left(c_2^a + \frac{c^b_2}{p^{1/2}} + \frac{c^c_2}{p}\,\right) 
+ q^4\,\cos\iota\,\left(c_3^a + \frac{c^b_3}{p^{1/2}} + \frac{c^c_3}{p}\,\right) \right.\nonumber \\ 
&+& \left. q\,\cos^2\iota\left(c^a_4 + \frac{c^b_4}{p^{1/2}} + \frac{c^c_4}{p}\,\right) 
+ q^3\,\cos^2\iota\,\left(c_5^a + \frac{c^b_5}{p^{1/2}} + \frac{c^c_5}{p}\,\right) 
+ q^2\,\cos^3\iota\,\left(c_6^a + \frac{c^b_6}{p^{1/2}} + \frac{c^c_6}{p}\,\right)
 \right.\nonumber \\ 
&+& \left. \left. q^4\,\cos^3\iota\,\left(c_7^a + \frac{c^b_7}{p^{1/2}} + \frac{c^c_7}{p}\,\right) 
+ q^3\,\cos^4\iota\,\left(c_8^a + \frac{c^b_8}{p^{1/2}} + \frac{c^c_8}{p}\,\right)
+ q^4\,\cos^5\iota\,\left(c_9^a + \frac{c^b_9}{p^{1/2}} + \frac{c^c_9}{p}\,\right) \right\} 
\right. \nonumber \\ 
&+& \left. \frac{q}{p^{7/2}}\,\cos\iota\,
\left\{ f^a_1 + \frac{f^b_1}{p^{1/2}} + q\,\left(f^a_2 + \frac{f^b_2}{p^{1/2}}\,\right) 
+ q^2\,\left(f^a_3 + \frac{f^b_3}{p^{1/2}}\,\right) 
+ \cos^2\iota \,\left(f^a_4 + \frac{f^b_4}{p^{1/2}}\,\right) \right. \right. \nonumber \\ 
&+& \left. \left.  q\,\cos^2\iota \,\left(f^a_5 + \frac{f^b_5}{p^{1/2}}\,\right) 
+ q^2\,\cos^2\iota\,\left(f^a_6 + \frac{f^b_6}{p^{1/2}}\,\right)\right\}\right ] \,.  \label{Ldotfit}
\end{eqnarray}
\end{widetext}
Similarly, a good fit to the evolution of $\iota$ is given by
\begin{widetext}
\begin{eqnarray}
(\dot{\iota})^{}_{\rm fit}  &=& \frac{32}{5} \frac{m^2_{\star}}{M^{}_{\bullet}} \,\frac{q\,\sin^2\iota}{\sqrt{Q}}\,
{p^{-5}}\,\left[ \,\frac{61}{24} + \frac{1}{p}\,\left(d_1^a + \frac{d^b_1}{p^{1/2}} + \frac{d^c_1}{p}\,\right) 
+ \frac{q^2}{p}\,\left(d_2^a + \frac{d^b_2}{p^{1/2}} + \frac{d^c_2}{p}\,\right)  
+ \frac{q}{p^{1/2}}\,\cos\iota\,\left(c_{10}^a + \frac{c^b_{10}}{p} + \frac{c^c_{10}}{p^{3/2}}\,\right)
\right. \nonumber \\ 
&+& \left. \frac{q^2}{p}\,\cos^2\iota\,\left(c_{11}^a + \frac{c^b_{11}}{p^{1/2}} + \frac{c^c_{11}}{p}\,\right)  
+ \frac{q^3}{p^{5/2}}\,\cos\iota\,\left\{ f^a_7 + \frac{f^b_7}{p^{1/2}} 
+ q\,\left(f^a_8 + \frac{f^b_8}{p^{1/2}}\,\right) + \cos^2\iota\,\left(f^a_9 + \frac{f^b_9}{p^{1/2}}\,\right)  
\right. \right. \nonumber \\ 
&+& \left. \left.  q\,\cos^2\iota\,\left(f^a_{10} + \frac{f^b_{10}}{p^{1/2}}\,\right) \right\}\right] 
\,,\label{idotfit}
\end{eqnarray}
\end{widetext}
where the values of the numerical fitting coefficients are:
\begin{widetext}
\begin{eqnarray}
&& d^a_1 = -10.7420\,, \qquad 
   d_1^b = 28.5942\,, \qquad 
   d^c_1 = -9.07738\,,\qquad 
   d^a_2 = -1.42836\,, \qquad 
   d_2^b = 10.7003\,, \nonumber \\[2mm] 
&& d^c_2 = -33.7090\,, \qquad 
   c^a_1 = -28.1517\,, \qquad 
   c^b_1 = 60.9607\,, \qquad 
   c^c_1 = 40.9998\,, \qquad 
   c^a_2 = -0.348161\,, \nonumber \\[2mm]
&& c^b_2 = 2.37258\,, \qquad  
   c^c_2 = -66.6584\,, \qquad 
   c^a_3 = -0.715392\,, \qquad 
   c^b_3 = 3.21593\,, \qquad 
   c^c_3 = 5.28888\,, \nonumber \\[2mm]  
&& c^a_4 = -7.61034\,, \qquad  
   c^b_4 = 128.878\,, \qquad 
   c^c_4 = -475.465\,, \qquad 
   c^a_5 = 12.2908\,, \qquad 
   c^b_5 = -113.125\,, \nonumber \\[2mm] 
&& c^c_5 = 306.119\,, \qquad  
   c^a_6 = 40.9259\,, \qquad 
   c^b_6 = -347.271\,, \qquad 
   c^c_6 = 886.503\,, \qquad 
   c^a_7 = -25.4831\,, \nonumber \\[2mm] 
&& c^b_7 = 224.227\,, \qquad  
   c^c_7 = -490.982\,, \qquad 
   c^a_8 = -9.00634\,, \qquad 
   c^b_8 = 91.1767\,, \qquad 
   c^c_8 = -297.002\,, \nonumber \\[2mm] 
&& c^a_9 = -0.645000\,, \qquad  
   c^b_9 = -5.13592\,, \qquad 
   c^c_9 = 47.1982\,, \qquad 
   f^a_1 = -283.955\,, \qquad 
   f^b_1 = 736.209\,, \nonumber \\[2mm] 
&& f^a_2 = 483.266\,, \qquad 
   f^b_2 = -1325.19\,, \qquad 
   f^a_3 = -219.224\,, \qquad 
   f^b_3 = 634.499\,, \qquad 
   f^a_4 = -25.8203\,, \nonumber \\[2mm] 
&& f^b_4 = 82.0780\,, \qquad 
   f^a_5 = 301.478\,, \qquad 
   f^b_5 = -904.161\,, \qquad 
   f^a_6 = -271.966\,, \qquad 
   f^b_6 = 827.319\,. 
\end{eqnarray}
\begin{eqnarray}
&& c_{10}^a = -0.0309341\,, \qquad 
   c_{10}^b = -22.2416\,, \qquad 
   c_{10}^c = 7.55265\,, \qquad 
   c_{11}^a = -3.33476\,, \qquad 
   c_{11}^b = 22.7013\,, \nonumber \\[2mm] 
&& c_{11}^c = -12.4700\,, \qquad  
   f^a_7 = -162.268\,, \qquad 
   f^b_7 = 247.168\,, \qquad 
   f^a_8 = 152.125\,, \qquad 
   f^b_8 = -182.165\,, \nonumber \\[2mm] 
&& f^a_9 = 184.465\,, \qquad 
   f^b_9 = -267.553\,, \qquad 
   f^a_{10} = -188.132\,, \qquad 
   f^b_{10} = 254.067\,.
\end{eqnarray}
\end{widetext}


%

\end{document}